\documentclass[12pt, notitlepage]{article}
\pdfoutput=1

\usepackage{float}
\usepackage{subfig}
\usepackage{graphicx}

\usepackage[T1]{fontenc}
\usepackage[utf8]{inputenc}
\usepackage{authblk}
\usepackage{amsmath,bm}
\usepackage[left=3cm,right=3cm,top=2cm,bottom=2cm]{geometry}

\renewcommand\footnotemark{}

\begin{document}


\title{Exploring Euclidean Dynamical Triangulations with a Non-trivial Measure Term}

\author[a,c]{D.N.~Coumbe}
                                                                                                                                                                                             
\author[a,b]{J.~Laiho}
\affil[a]{\emph{SUPA, School of Physics and Astronomy, University of Glasgow, Glasgow, G12 8QQ, UK}}
\affil[b]{\emph{Department of Physics, Syracuse University, Syracuse, NY 13244, USA}}
\affil[c]{\emph{Institute of Physics, Jagellonian University, Krakow, PL 30-059, Poland}\footnote{E-mail: d.coumbe@physics.gla.ac.uk, jwlaiho@syr.edu}}
\date{\today}
\maketitle


\begin{abstract}

We investigate a nonperturbative formulation of quantum gravity defined via Euclidean dynamical triangulations (EDT) with a non-trivial measure term in the path integral. We are motivated to revisit this older formulation of dynamical triangulations by hints from renormalization group approaches that gravity may be asymptotically safe and by the emergence of a semiclassical phase in causal dynamical triangulations (CDT). 

We study the phase diagram of this model and identify the two phases that are well known from previous work: the branched polymer phase and the collapsed phase. We verify that the order of the phase transition dividing the branched polymer phase from the collapsed phase is almost certainly first-order. The nontrivial measure term enlarges the phase diagram, allowing us to explore a region of the phase diagram that has been dubbed the crinkled region. Although the collapsed and branched polymer phases have been studied extensively in the literature, the crinkled region has not received the same scrutiny. We find that the crinkled region is likely a part of the collapsed phase with particularly large finite-size effects. Intriguingly, the behavior of the spectral dimension in the crinkled region at small volumes is similar to that of CDT, as first reported in arXiv:1104.5505, but for sufficiently large volumes the crinkled region does not appear to have 4-dimensional semiclassical features. Thus, we find that the crinkled region of the EDT formulation does not share the good features of the extended phase of CDT, as we first suggested in arXiv:1104.5505.  This agrees with the recent results of arXiv:1307.2270, in which the authors used a somewhat different discretization of EDT from the one presented here. 

\end{abstract}
\clearpage


\begin{section}{Introduction}

Formulating a consistent, predictive theory of quantum gravity is one of the great outstanding challenges of theoretical physics. 
Although it is possible to formulate gravity as an effective field theory at low energies  \cite{Donoghue:1997hx}, a perturbative treatment of general relativity is nonrenormalizable by power counting; counterterms that cannot be removed by the equations of motion have been found explicitly at the one-loop level when including matter content \cite{'tHooft:1974bx} and at the two-loop level for pure gravity \cite{Goroff:1985th}. However, it is possible that gravity is nonperturbatively renormalizable. The conjecture that this could be the case was made by Weinberg in Ref.~\cite{Weinberg79}, where he coined the term asymptotic safety to describe this scenario. If gravity were asymptotically safe, it would be ultraviolet-complete and describable in terms of a finite number of free parameters, making it effectively renormalizable when formulated nonperturbatively. The renormalization group flow of couplings would end at a non-trivial fixed point in the ultraviolet, with a finite-dimensional critical surface. The dimensionless couplings at such a fixed point are not necessarily small, so that perturbation theory may not be a reliable guide to whether the theory is actually renormalizable. 

The lattice is a powerful tool for performing nonperturbative calculations in quantum field theory and is the standard for precision calculations of quantum chromodynamics in the nonperturbative regime \cite{Beringer:1900zz,Colangelo:2010et}. In a lattice formulation of an asymptotically safe field theory, the fixed point would appear as a second-order critical point, the approach to which would define a continuum limit. The divergent correlation length characteristic of a second-order phase transition would allow one to take the lattice spacing to zero while keeping observable quantities fixed in physical units.

Evidence for the existence of a fixed point in the 4d theory of quantum gravity without matter has come mainly from continuum functional renormalization group methods
\cite{Reuter:2001ag,Lauscher:2001ya,Litim:2003vp,Codello:2007bd,Codello:2008vh,Benedetti:2009rx}, and lattice calculations \cite{Ambjorn05,Ambjorn:1998xu,Ambjorn:2008wc}.
The renormalization group methods have provided evidence for the existence of a non-trivial fixed point in a variety of truncations of the renormalization group equations, indicating that the dimension of the ultraviolet critical surface is finite. Furthermore, truncations using more than three independent couplings indicate that the critical surface remains three-dimensional \cite{Codello:2007bd,Codello:2008vh,Benedetti:2009rx}. Given that the truncation of the effective action makes it difficult to systematically assess the reliability of the results obtained using this approach, a method that allows calculations with controlled systematic errors is desirable. In principle this is possible using lattice methods.

One of the original attempts to formulate quantum gravity on the lattice is Euclidean dynamical triangulations (EDT) \cite{Ambjorn:1991pq,Agishtein:1991cv}, 
which defines a spacetime comprised of locally flat $n$-simplices of fixed edge length, 
and the dynamics is contained in the connectivity of the simplices. 
Early studies of EDT showed that there are two phases, neither of which resembles four-dimensional semi-classical general relativity. The two phases in the original EDT model were at least initially thought to be separated by a second-order phase transition \cite{Ambjorn:1991pq,Catterall:1994pg,Ambjorn:1995dj}. However, upon a more detailed analysis, which included larger lattice volumes, the phase transition was revealed to be first-order \cite{Bialas:1996wu,deBakker:1996zx}\footnote{See also the recent work of Ref.~\cite{Rindlisbacher:2013gka}}, making it unlikely that a continuum limit exists, at least in the simplest implementation of the model. 

The difficulties encountered in the original EDT model led Ambjorn and Loll \cite{Ambjorn:1998xu} to introduce a causality constraint on the set of 
triangulations over which the path integral is taken. 
The method of causal dynamical triangulations (CDT) distinguishes between space-like and time-like links on the lattice so that an explicit foliation 
of the lattice into space-like hypersurfaces of fixed topology (usually chosen to be $S^{3}$) can be introduced. Work on this formulation has produced a number of interesting and promising results. In addition to the emergence of a 4-dimensional semiclassical phase resembling de Sitter space \cite{Ambjorn:1998xu,Ambjorn:2008wc}, 
it is found that for CDT the fractal (spectral) dimension runs as a function of the distance scale probed.

Our motivation for revisiting the original Euclidean theory is as follows. Firstly, the CDT restriction to a fixed foliation is potentially at odds with general covariance, 
though it may amount to a choice of gauge \cite{Markopoulou:2004jz}. Recent work suggests that it is possible to relax the restriction to a fixed foliation \cite{Jordan:2013iaa}, though a different treatment of space-like and time-like links is still necessary. It is interesting to establish whether or not the same results can be obtained 
using an explicitly space and time symmetric EDT approach. Secondly, renormalization group studies \cite{Codello:2008vh,Benedetti:2009rx} suggest that the ultraviolet critical surface may be 3-dimensional, so it seems 
worthwhile to revisit EDT with additional parameters in the bare lattice action \cite{Bruegmann:1992jk,Bilke:1998vj}. Both the CDT and EDT models 
include the bare Newton and cosmological constants, but CDT also includes a third parameter, the ratio of the lengths of space-like and time-like links on the lattice, $\Delta$. The parameter $\Delta$ enlarges the phase diagram, and increasing $\Delta$ away from zero permits the existence of the de Sitter phase.  
Thus, we investigate the result of including a third parameter in the bare lattice action of EDT \cite{Laiho:2011ya,Laiho2014,Coumbe:2012qr}. In this work we add a local measure term to the calculation. Such a term has been considered in earlier works, where it was tentatively concluded that a new phase, dubbed the crinkled phase, appears \cite{Bilke:1998vj,Thorleifsson:1998qi}. This region of the EDT phase diagram had not been studied in detail in 4-dimensions, and in light of recent work that supports the asymptotic safety scenario for gravity, we are taking a closer look. We found in our previous work \cite{Laiho:2011ya} that there exists a region of the phase diagram (the crinkled region) where EDT calculations give results for the spectral dimension that are strikingly similar to those of CDT. However, we find that with improved statistics and larger volumes that the region of the phase diagram studied in \cite{Laiho:2011ya} does not have 4-dimensional semiclassical behavior.  A similar study was recently presented in Ref. \cite{Ambjorn:2013eha}.

The layout of this paper is as follows. In Sec. 2 we discuss the numerical implementation of our simulations. In Sec. 3 the phase diagram of EDT is given, along with the analysis that leads to this picture. In Sec. 4 we focus on the crinkled region of EDT. Finally, a brief discussion and some concluding remarks are given.

\end{section}

  
\begin{section}{Numerical Implementation of EDT}
\label{Num}

EDT is an attempt to formulate quantum gravity using lattice methods.  
EDT was originally formulated in two-dimensions as a nonperturbative regularization of bosonic string theory \cite{Ambjorn:2011cg}. 
This two dimensional approach to gravity coupled to conformal matter was shown to correspond to non-critical bosonic string theory \cite{Ambjorn:2002nt}. 
Results from lattice calculations agree with continuum calculations in non-critical string theory wherever the two have been compared \cite{Ambjorn:2002nt}. 
Motivated by the successes of the two dimensional theory, EDT was generalised to three \cite{Ambjorn:1991wq,Agishtein:1991ta,Boulatov:1991hg} and four dimensions 
\cite{Ambjorn:1991pq,Agishtein:1991cv}. In this work we consider EDT in four dimensions only.

Analytical methods in 4-dimensional dynamical triangulations have thus far proved intractable due to the difficult nonperturbative sum over geometries. 
However, with the advent of powerful computational tools, numerical methods can now be successfully employed. In this section we review the numerical implementation of dynamical triangulations.

In EDT the path integral for Euclidean gravity is given by the discrete partition function

\begin{equation} \label{eq:EDTPartitionFunction}
Z_{E}={\sum_{T}}\frac{1}{C_{T}}\left[\prod_{j=1}^{N_{2}}\mathcal{O}\left(t_{j}\right)^{\beta}\right]e^{-S_{E}}
\end{equation}

\noindent where $C_{T}$ is a symmetry factor that divides out the number of equivalent ways of labelling the vertices in the triangulation $T$. The term in brackets is the nontrivial local measure factor. The Euclidean Einstein-Regge action is

\begin{equation} \label{eq:DiscreteEHAction}
S_{E}=-\kappa_{2}N_{2}+\kappa_{4}N_{4},
\end{equation}

\noindent where $N_{i}$ is the number of simplices of dimension $i$, and \begin{math}\kappa_{2}\end{math} and \begin{math}\kappa_{4}\end{math} are related to the 
bare Newton's constant $G_{N}$ and the bare cosmological constant $\Lambda$. The product in Eq. (\ref{eq:EDTPartitionFunction}) is over all 2-simplices (triangles) $t$, and \begin{math}\mathcal{O}\left(t_{j}\right)\end{math} is the order of the triangle $t_j$, i.e. the number of four-simplices to which the triangle belongs. 
The non-trivial measure term corresponds in the continuum to a nonuniform weighting of the measure by 
$\left[\mbox{det}\left(-g\right)\right]^{\beta/2}$ \cite{Bruegmann:1992jk}. We vary \begin{math}\beta\end{math} as an additional independent parameter in the bare 
lattice action (after exponentiating the measure term). Most of the previous work on EDT considered 
the partition function with $\beta=0$ only.

The simple form of Eq.~(\ref{eq:DiscreteEHAction}) for the discrete Einstein-Regge action $S_{E}$ arises as follows. 
\noindent The discrete Euclidean-Regge action is \cite{Regge:1961px}

\begin{equation} \label{eq:GeneralEinstein-ReggeAction}
S_{E}=-\kappa\sum V_{2}\left(2\pi-\sum\theta\right)+\lambda\sum V_{4},
\end{equation}

\noindent where $\kappa=\left(8\pi G_{N} \right)^{-1}$, $\lambda=\kappa\Lambda$, and $\theta=\rm{arccos}\left(\frac{1}{4}\right)$. The volume of a $d$-simplex is 

\begin{equation} \label{eq:SimplexVolume}
V_{d}=\frac{\sqrt{d+1}}{d!\sqrt{2^{d}}}.
\end{equation}

\noindent Rewriting Eq.~(\ref{eq:GeneralEinstein-ReggeAction}) in terms of the bulk 
variables $N_{i}$, and using Eq.~(\ref{eq:SimplexVolume}) for the simplicial volume, one finds

\begin{equation}\label{eq:DiscAction}
S_{E}=-\frac{\sqrt{3}}{2}\pi\kappa N_{2}+N_{4}\left(\kappa\frac{5\sqrt{3}}{2}\mbox{arccos}\frac{1}{4}+\frac{\sqrt{5}}{96}\lambda\right).
\end{equation}

\noindent Setting $\kappa_{2}=\frac{\sqrt{3}}{2}\pi\kappa$ and $\kappa_{4}=\kappa\frac{5\sqrt{3}}{2}\mbox{arccos}\left(\frac{1}{4}\right)+\frac{\sqrt{5}}{96}\lambda$ 
one obtains Eq. (\ref{eq:DiscreteEHAction}) for the discrete Einstein-Regge action in terms of the bulk variables $N_{2}$ and $N_{4}$.

 A $d$-dimensional simplicial manifold is constructed by gluing $d$-simplices together along their $(d-1)$-dimensional faces. 
To each $d$-simplex there exists a simplex label and a set of combinatorially unique $(d+1)$ vertex labels. The set of combinatorial triangulations was used in most early simulations of EDT. 
However, the constraint of combinatorial uniqueness can be relaxed to include a larger set of degenerate triangulations in which the neighbours of a given simplex are 
no longer unique \cite{Bilke:1998bn}. Triangulations in this new set do not satisfy the combinatorial manifold constraints. It has been shown numerically that simulations using degenerate triangulations lead to the same phase structure as combinatorial triangulations for the well-studied case of $\beta=0$ in 4-dimensions, and also in the case of 3-dimensions \cite{Bilke:1998bn,Thorleifsson:1998nb}. In both instances, using the set of degenerate triangulations leads to a factor of \begin{math}\sim\end{math}10 
reduction in finite-size effects compared to combinatorial triangulations \cite{Bilke:1998bn}, with similar results. Thus, in this work we adopt the set of degenerate triangulations. The recent work of Ref.~\cite{Ambjorn:2013eha} used the set of combinatorial triangulations, so our work provides a cross-check of their results with a different discretization.

We have made various checks of our code against the literature for both combinatorial 
triangulations \cite{deBakker:1996zx}, and degenerate triangulations \cite{Bilke:1998bn}, and good agreement has been found. For combinatorial triangulations we have compared our calculation of several different quantities against the literature. The susceptibility of the scalar curvature and the number of simplices within a varying geodesic distance from a randomly chosen origin were computed for three different values of the bare parameters and compared with the work of Ref.~\cite{deBakker:1996zx}, and good agreement was found. We also compared our calculation of a third quantity, the node order susceptibility, with the work of Ref.~\cite{Bilke:1998vj} for values of $\beta\neq 0$. Good agreement was found for this quantity for all ten values of the bare parameters in the comparison. For degenerate triangulations we compared our critical value of $\kappa_{2}$ with that of Ref. \cite{Bilke:1998bn} for $\beta=0$. Again, good agreement was found. 

The partition function of Eq.~(\ref{eq:EDTPartitionFunction}) is evaluated using Monte Carlo methods. An ensemble of 4-dimensional degenerate triangulations with 
fixed topology $S^{4}$ is generated with the Boltzmann weight of Eq.(\ref{eq:EDTPartitionFunction}) using the Metropolis algorithm and a set of five local update moves. These update moves are known as the Pachner moves, and are by now standard \cite{Pachner:1991:PHM:107892.107898}. For spacetime dimension $d\leq4$ the Pachner moves are known to be ergodic for combinatorial and degenerate triangulations, i.e. one can go from any triangulation to any other via a repeated application of moves from the set. The Pachner moves \cite{Pachner:1991:PHM:107892.107898} are a variant of the Alexander moves \cite{Alexander:1930aa}, and consist of the following five operations: inserting or removing a vertex, replacing a tetrahedron by an edge or vice-versa, or flipping a triangle (a move that is its own inverse). The update moves are called in random order and it is 
ensured that the number of accepted moves of each type is approximately equal. We define a sweep to be $10^{8}$ attempted moves, with the computational time for one sweep exhibiting a very small volume dependence. 

In four dimensions the Pachner moves are only ergodic if the number of simplices is allowed to vary. However, it is convenient to keep $N_{4}$ approximately fixed by the inclusion 
of a term $\delta\lambda|N_{4}-V|$ in the action. This term permits fluctuations in $N_{4}$ of magnitude 
$\delta N_{4}=\left(\left\langle N_{4}^{2} \right\rangle^{2} - \left\langle N_{4} \right\rangle^{2}\right)^{1/2}=\left(\frac{1}{2\delta\lambda}\right)^{1/2}$. For most of our runs we have set $\delta\lambda$=0.04. 
As usual in dynamical triangulations, the bare cosmological constant, or equivalently $\kappa_{4}$, must be tuned to its critical value so that an infinite volume limit can be taken \cite{deBakker:1994zf}. 
This leaves a two dimensional parameter space, which is explored by varying $\kappa_{2}$ and $\beta$.

It is important to make sure that a run has thermalized, especially when exploring a region of the phase diagram that might have particularly long auto-correlation lengths. We introduce an observable known as the volume-volume correlator $c_{N_{4}}(x)$ that depends on the long distance structure of the geometry, and whose behavior we have found is a good test of whether or not the ensemble is thermalized. To calculate the volume-volume correlator one begins by defining a randomly chosen vertex from the ensemble of triangulations to be the origin, $O$. One then moves radially outwards from this point by hopping to an adjacent vertex. We can thus define a geodesic distance $\tau$ from the origin $O$, counting the number of simplices within a shell of radius $\tau$. We call the total number of 4-simplices in a spherical shell a geodesic distance \begin{math}\tau\end{math} from the origin $N_{4}^{\rm shell}(\tau)$. 

The volume-volume correlator is defined similarly to the one introduced in Ref. \cite{Ambjorn05} to study CDT,

\begin{equation}\label{volvolcorr}
C_{N_{4}}\left(\delta\right)=\sum_{\tau=1}^t\frac{\left\langle N_{4}^{\rm shell}(\tau)N_{4}^{\rm shell}(\tau+\delta)\right\rangle }{N_{4}^{2}}.
\end{equation}

\noindent \begin{math}N_{4}\end{math} is the total number of 4-simplices and the normalization of the correlator is chosen following Ref.~\cite{Ambjorn05} such that $\sum_{\delta=0}^{t-1}C_{N_{4}}\left(\delta\right)=1.$ We introduce the rescaled variables $x$ and \begin{math}c_{N_{4}}\left(x\right)\end{math} such that \begin{math}x=\delta/N_{4}^{1/D_{H}}\end{math} and 

\begin{equation}\label{cvol}
c_{N_{4}}\left(x\right)=N_{4}^{1/D_{H}}C_{N_{4}}\left(N_{4}^{1/D_{H}}x\right),
\end{equation} 

\noindent where $D_{H}$ is the Hausdorff dimension. If the Hausdorff dimension \begin{math}D_{H}\end{math} is chosen correctly, \begin{math}c_{N_{4}}\left( x \right)\end{math} will be invariant under a change in four-volume \begin{math}N_{4}\end{math}.

Figure~\ref{Therm8KB02} shows a typical plot of the peak of the volume-volume correlator $c_{N_{4}}(x)$ as a function of Monte Carlo time. Figure~\ref{ThermFit8KB02} is a zoomed-in version of Fig. \ref{Therm8KB02}, focusing on the configuration range we believe is thermalized. A comparison between the first half and the second half of the data set shows statistical agreement.  All of the statistical errors in this paper are computed using a single-elimination (binned) jackknife procedure, after blocking the data as necessary to account for autocorrelation errors.  The statistical errors increase with increasing block size when autocorrelation errors are important; the errors do not change with increasing block size when autocorrelation errors are small.  We choose the block size of our observables so that the statistical errors are not dramatically underestimated.

\begin{figure}[H]
  \centering
  \includegraphics[width=0.8\linewidth,natwidth=610,natheight=642]{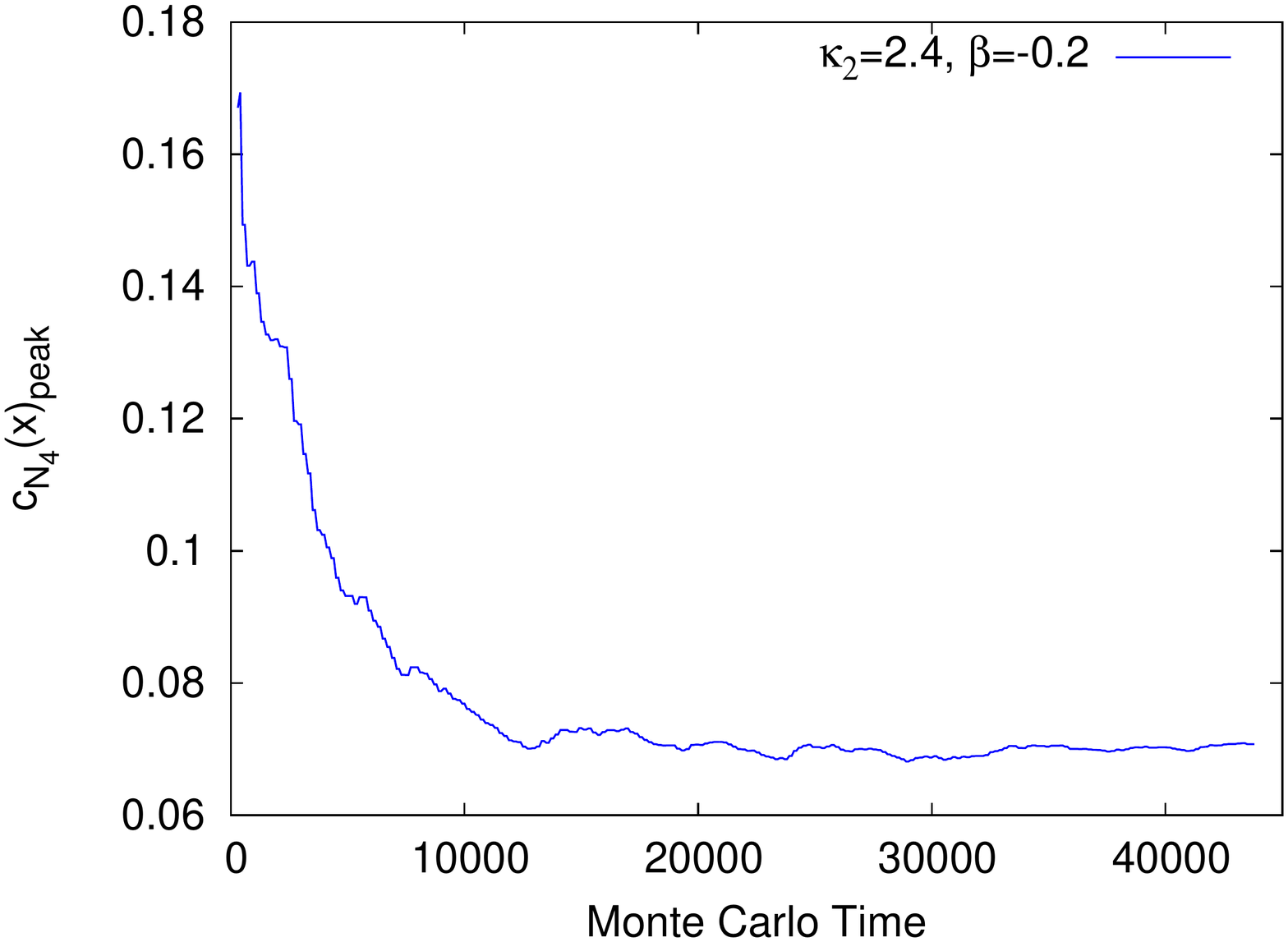}
  \caption{\small A thermalization check as performed via a plot of the peak in the volume-volume correlator $c_{N_{4}}(x)$ as a function of Monte Carlo time (in units of $10^{8}$ attempted moves).}
\label{Therm8KB02}
\end{figure}

\begin{figure}[H]
  \centering
  \includegraphics[width=0.8\linewidth,natwidth=610,natheight=642]{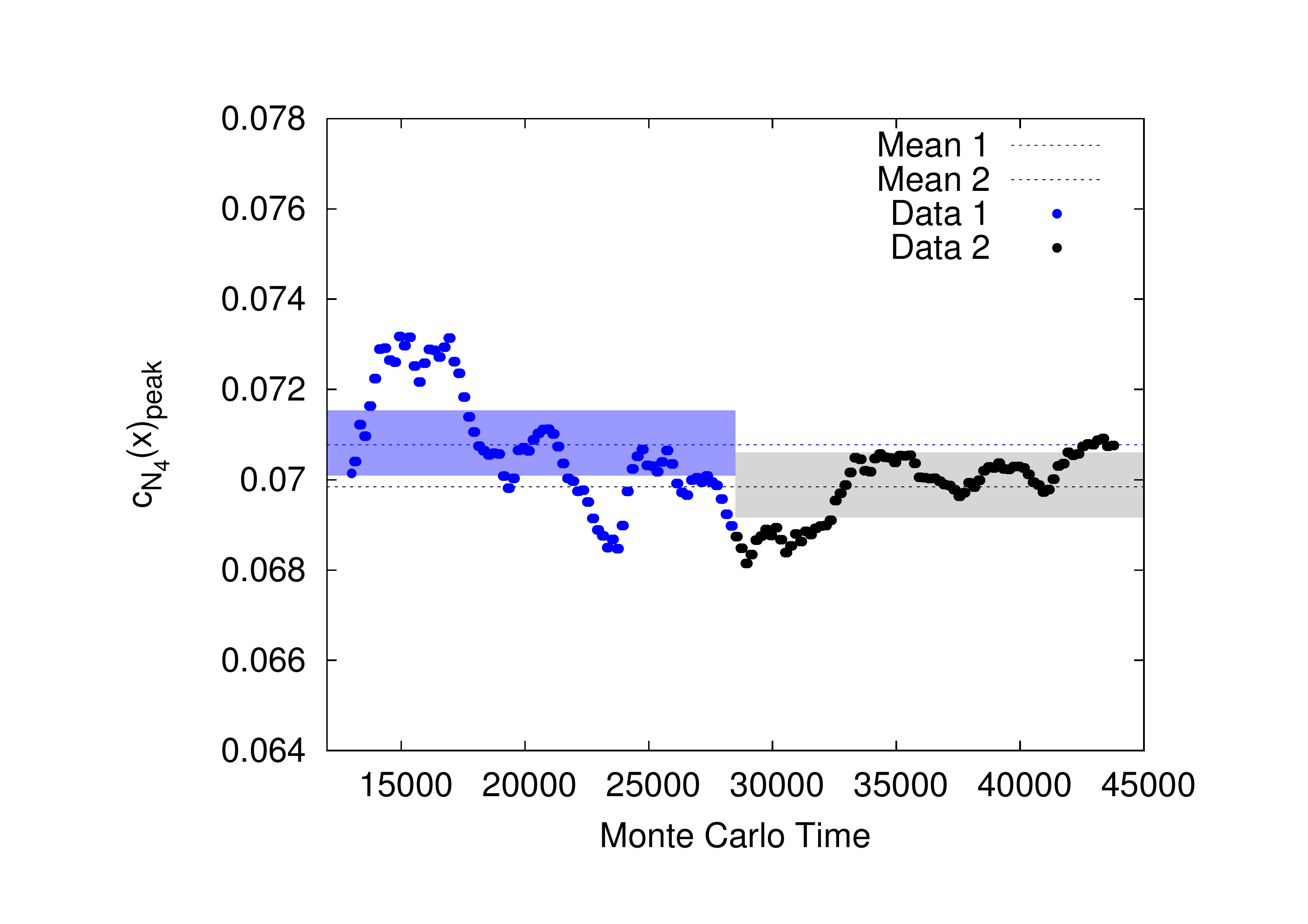}
  \caption{\small A thermalization check as performed via a plot of the peak in the volume-volume correlator $c_{N_{4}}(x)$ as a function of Monte Carlo time (in units of $10^{8}$ attempted moves). The data range we believe to be thermalized is divided in to two data sets that are compared with each other for statistical agreement.}
\label{ThermFit8KB02}
\end{figure}

The different types of fractal dimension are useful in characterizing the lattices that emerge from the simulations. The Hausdorff dimension is one such generalization of the concept of dimension to fractal geometries and non-integer values. The Hausdorff dimension $D_{H}$ is calculated by determining how the radius of a sphere $r$ scales with volume $V$ in the limit that $r\rightarrow 0$, and is given by

\begin{equation}
D_{H}=\lim_{r \to 0} \frac{\rm{log} \left(\emph{V} \left(\emph{r} \right)\right)}{\rm{log} \left(\emph{r} \right)}.
\end{equation}

Another measure of the fractal dimension of a space is the spectral dimension $D_{S}$. The spectral dimension $D_{S}$ is related to the return probability $P_{r}\left(\sigma\right)$ of a random walk over the geometry after $\sigma$ diffusion steps. It is defined by,

\begin{equation}
D_{S}=-2\frac{d\rm{log}\langle \emph{P}_{\emph{r}}\left(\sigma\right)\rangle}{d\rm{log}\sigma}.
\label{spec2}
\end{equation}

\end{section}


\begin{section}{The Phase Diagram of EDT}
\begin{subsection}{Overview}

This section explores the phase diagram of EDT with a non-trivial measure term in the path integral. The parameter $\kappa_{4}$ is adjusted to take the infinite volume limit \cite{DeBakker:1994az}, leaving a two-dimensional parameter space that can be explored by varying $\kappa_2$ and $\beta$. 

The parameter space of EDT is enlarged via the inclusion of the new parameter $\beta$ that is associated with the non-trivial measure term of Eq.~(\ref{eq:DiscreteEHAction}). The result is a phase diagram with three regions; the branched polymer phase, the collapsed phase, and the crinkled region.  We begin this section by showing that we are able to reproduce the expected values of the fractal dimensions in the branched polymer phase and in the collapsed phase, which have been thoroughly investigated in the literature \cite{Bilke:1998bn, deBakker:1994zf,  Ambjorn:1999nc}.  This gives us confidence that our subsequent investigations of the fractal dimensions in the crinkled region of the phase diagram are reliable.  We then present our study of the phase diagram of the EDT model with a non-trivial measure term.  We find evidence that the branched polymer phase and the collapsed phase are separated by a first-order phase transition line AB, as shown in Fig.~\ref{PhaseDiagram}, though for sufficiently large negative $\beta$ and large $\kappa_2$ it is difficult to determine the location and order of the phase transition reliably due to a decrease in the Monte Carlo acceptance rate.  We also find that the line CD separating the collapsed phase from the crinkled region is a softer transition that is consistent with an analytic crossover.\interfootnotelinepenalty=10000 \footnote{\scriptsize The transition could be third or higher order, since it is very difficult to distinguish an analytic cross-over from a higher-order transition using numerical methods.}  

\begin{figure}[H] 
  \centering
  \includegraphics[width=0.6\linewidth,natwidth=610,natheight=642]{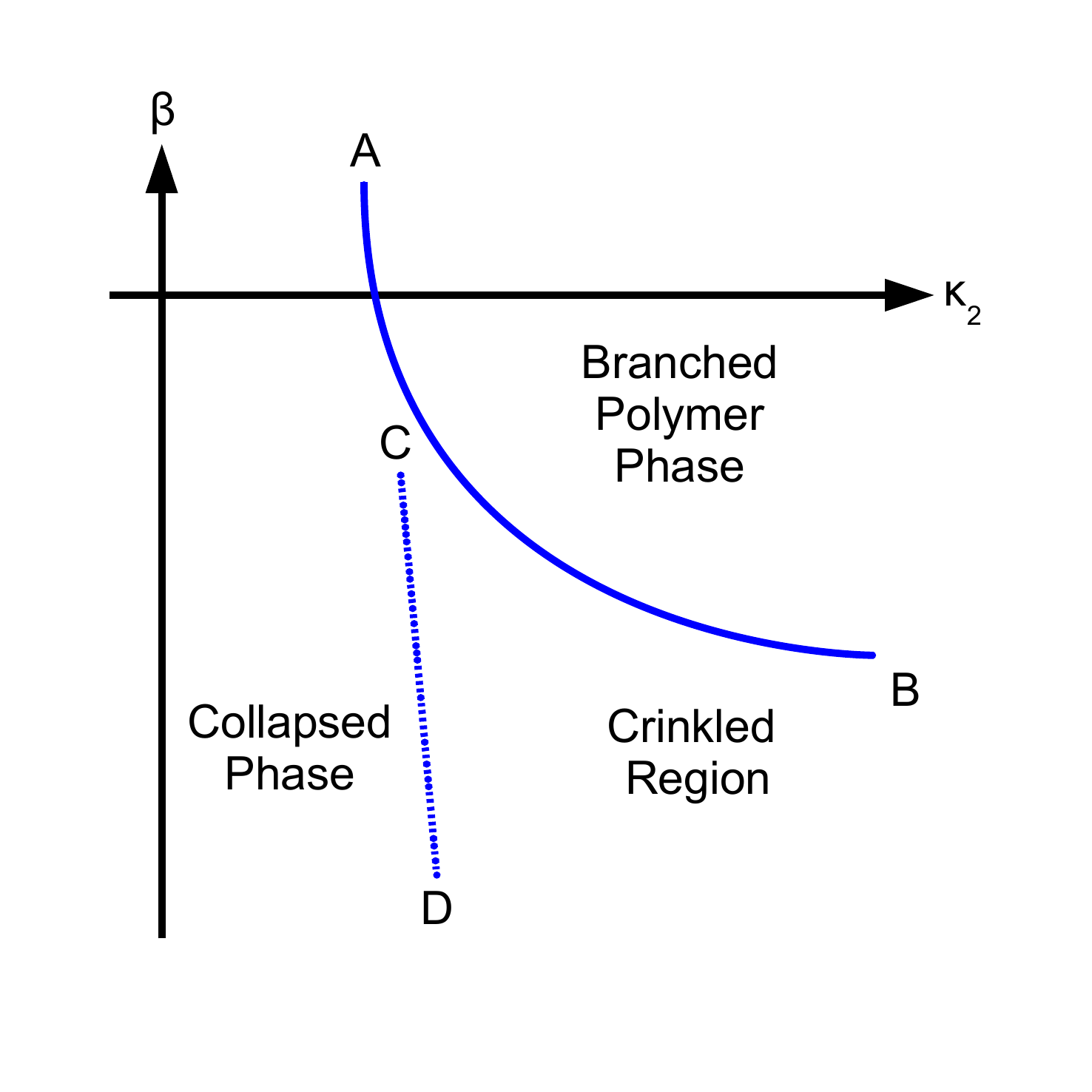}
  \caption{\small A schematic of the phase diagram of EDT with a non-trivial measure term.}
\label{PhaseDiagram}
\end{figure}  

The model is in the collapsed phase for sufficiently small values of $\kappa_{2}$. The collapsed phase is characterised by a very large, and possibly infinite, fractal Hausdorff dimension $D_{H}$. In this phase the spectral dimension $D_{s}$ also becomes very large, and possibly infinite in the infinite volume limit. In the collapsed phase there are a small number of highly connected vertices, so that a large number of simplices share a few common vertices. Thus, in this phase the vertices have a large coordination number. The volume distribution within the collapsed phase, as measured by the number of simplices within a given geodesic distance, is not well-described by Euclidean de Sitter space in four dimensions. The geometric properties of the collapsed phase suggest that it is unphysical.

\begin{figure}[H] 
  \centering
  \includegraphics[width=0.8\linewidth,natwidth=610,natheight=642]{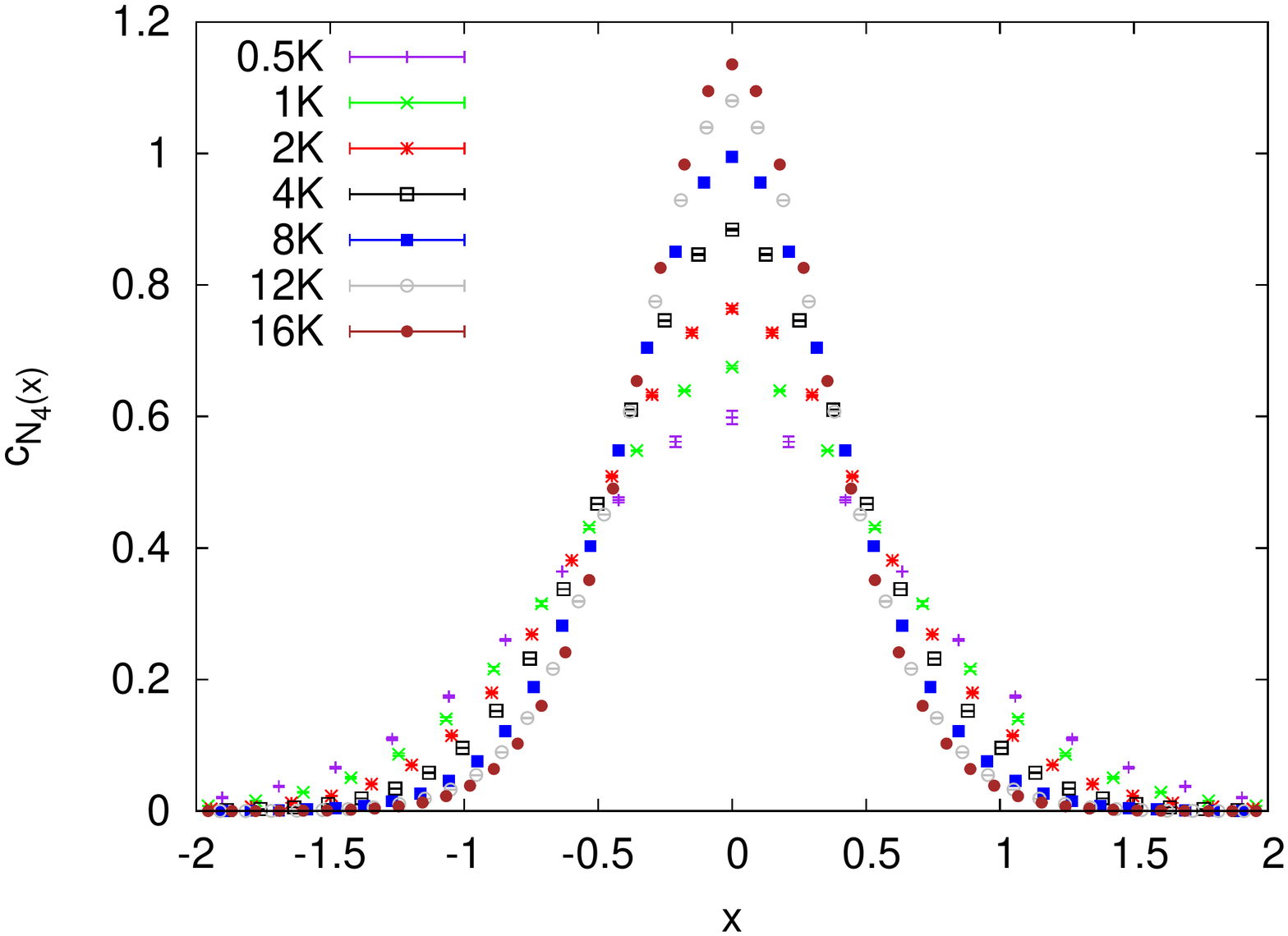}
  \caption{\small The volume-volume correlator $c_{N_{4}}(x)$ as a function of $x$ in the collapsed phase ($\kappa_{2}=1.0$, $\beta=0$) for several different lattice volumes. The value $D_{H}=4$ was chosen for the rescaling. The disagreement between the curves shows that $D_{H}\neq4$ in this phase.}
\label{VolCorr4K8KColl}
\end{figure}

\begin{figure}[H]
  \centering
  \includegraphics[width=0.8\linewidth,natwidth=610,natheight=642]{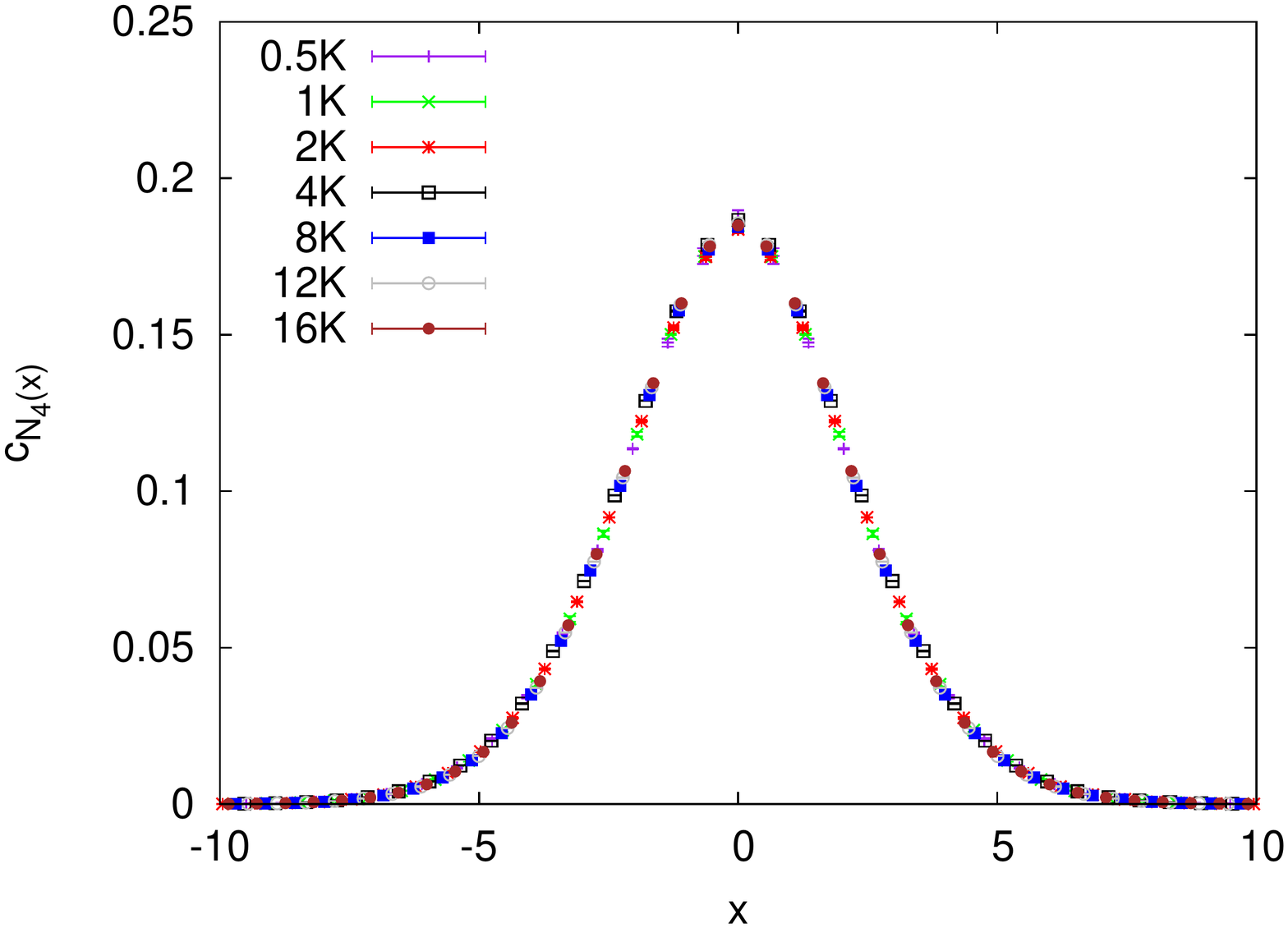}
  \caption{\small The volume-volume correlator $c_{N_{4}}(x)$ as a function of $x$ in the collapsed phase ($\kappa_{2}=1.0$, $\beta=0$) for several different lattice volumes and for a rescaling dimension of $D_{H}=16$.}
\label{VolCorr4K8KColl_v2}
\end{figure}

\begin{figure}[H]
  \centering
  \includegraphics[width=0.8\textwidth,natwidth=610,natheight=642]{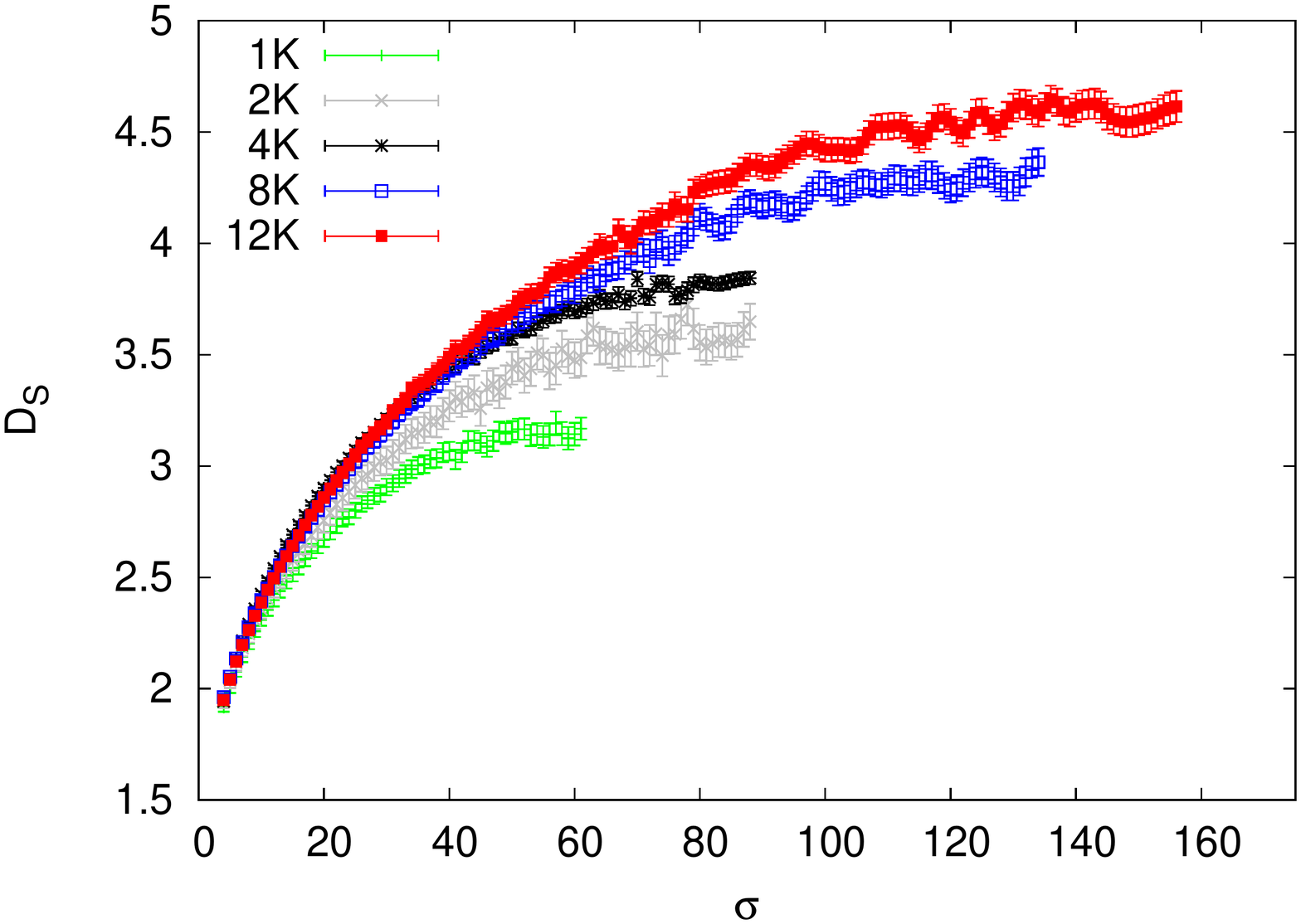}
  \caption{\small The spectral dimension $D_{S}$ as a function of diffusion time $\sigma$ in the collapsed phase ($\kappa_{2}=1.0$, $\beta=0$) for several different lattice volumes.
\label{fig:SpecCol}}
\end{figure}

Figure~\ref{VolCorr4K8KColl} is a plot of the rescaled volume-volume correlator $c_{N_{4}}(x)$, as a function of the rescaled time variable $x$ in the collapsed phase for seven different lattice volumes. In Fig.~\ref{VolCorr4K8KColl} $c_{N_{4}}(x)$ is calculated assuming a scaling dimension $D_{H}=4$. The quantity $c_{N_{4}}(x)$ should be volume independent if the rescaling is done with the correct value of the Hausdorff dimension. However, as can be seen in Fig. \ref{VolCorr4K8KColl} the curves for $c_{N_{4}}(x)$ disagree when $D_{H}=4$.  Figure \ref{VolCorr4K8KColl_v2} shows the correlator rescaled assuming $D_{H}=16$; at this value the agreement between the rescaled curves is good, demonstrating that the Hausdorff dimension is considerably greater than 4 in the collapsed phase, as expected.     

The spectral dimension in the collapsed phase is shown in Fig.~\ref{fig:SpecCol}. One can see from Fig.~\ref{fig:SpecCol} that the spectral dimension increases beyond $D_{S}=4$ as the lattice volume is increased. When the lattice volume is bigger it is possible to follow $D_{S}(\sigma)$ out to larger values of $\sigma$ before finite-volume effects set in and cause $D_{S}(\sigma)$ to decrease.

The model is in the branched polymer phase for sufficiently large values of $\kappa_{2}$ and $\beta$. Within this phase the geometry of triangulations undergoes numerous instances of ``pinching'' in which the geometry collapses to a minimal neck and branches off into polymer-like baby universes \cite{Catterall:1995aj}. This phase has a highly irregular geometry, exhibiting a fractal tree-like structure even on large scales. The Hausdorff dimension of a branched polymer is expected to be $D_{H}=2$, a result that is confirmed by our calculations, as illustrated in Fig. \ref{VolCorr4K8KBran}. Figure \ref{VolCorr4K8KBran} is a plot of the rescaled volume-volume correlator $c_{N_{4}}(x)$ as a function of $x$ in the branched polymer phase, for four different lattice volumes 8K, 4K, 2K, and 1K. Here $c_{N_{4}}(x)$ is rescaled assuming $D_{H}=2$, and good agreement is found over the entire range of $x$ values.

The spectral dimension of a branched polymer with $D_{H}$=2 is expected to be $D_{s}=4/3$ \cite{Jonsson:1997gk,Ambjorn:1997jf}, and this is confirmed by our numerical calculations, as can be seen in Fig.~\ref{fig:SpecBP}. The spectral dimension is calculated in the branched polymer phase for four different lattice volumes. It is determined by taking an average over a range of $\sigma$ values from $\sigma=200$-$450$, where a good plateau is seen and discretization effects appear to be negligible. The results are presented in Tab.~\ref{BPTable}. As can be seen in Tab.~\ref{BPTable}, the spectral dimension slightly undershoots 4/3 for smaller volumes, but becomes consistent with 4/3 for volumes of approximately 4000 simplices or larger.

\begin{figure}[H] 
  \centering
  \includegraphics[width=0.8\linewidth,natwidth=610,natheight=642]{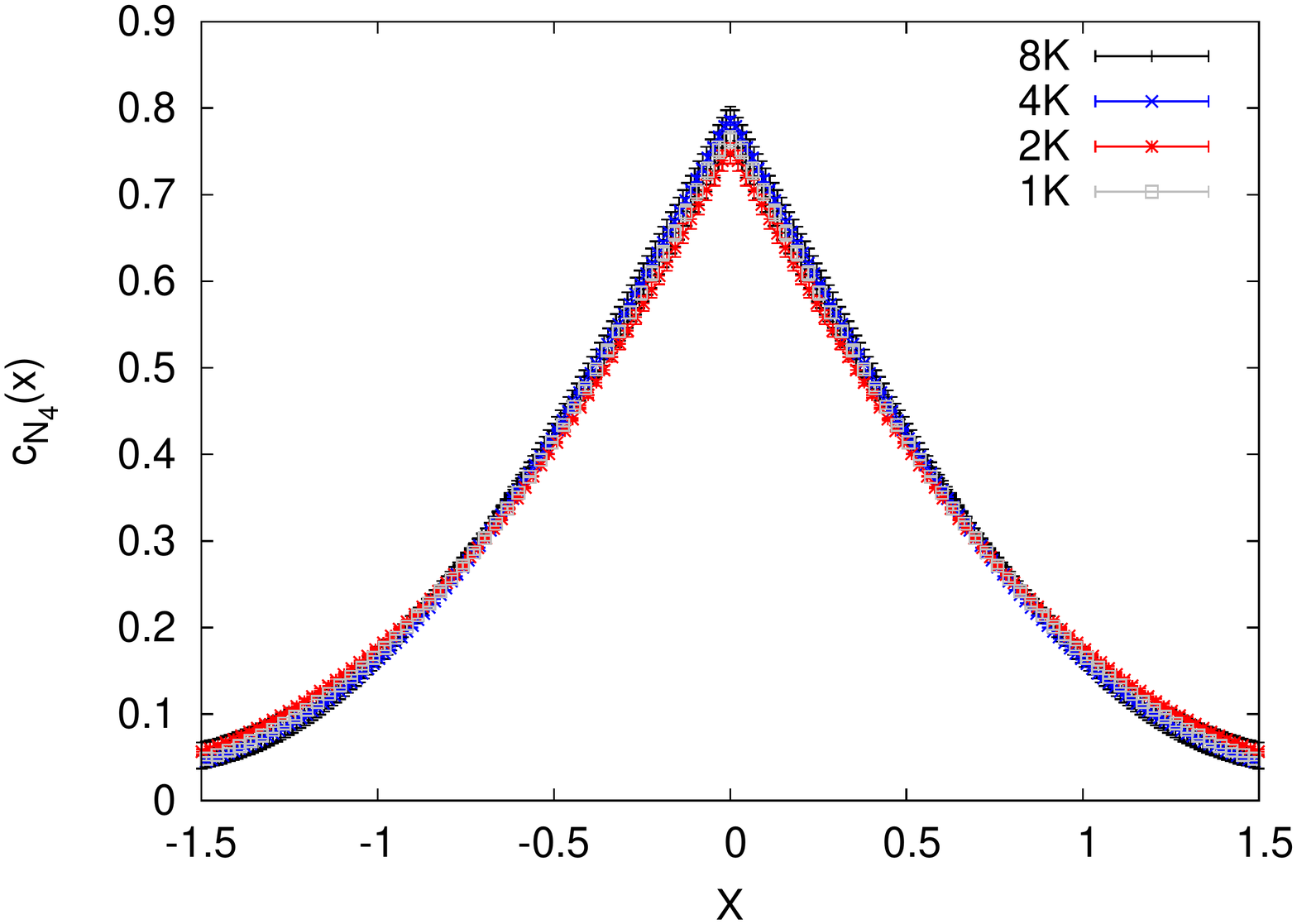}
  \caption{\small The volume-volume correlator $c_{N_{4}}(x)$ as a function of $x$ in the branched polymer phase for 4 different volumes 8K, 4K, 2K, 1K. When rescaling the volume-volume correlator we assume a Hausdorff dimension of $D_{H}=2$.}
\label{VolCorr4K8KBran}
\end{figure}

\begin{figure}[H]\label{fig:SpecBP}
\subfloat[]{\includegraphics[width=0.55\textwidth,natwidth=610,natheight=642]{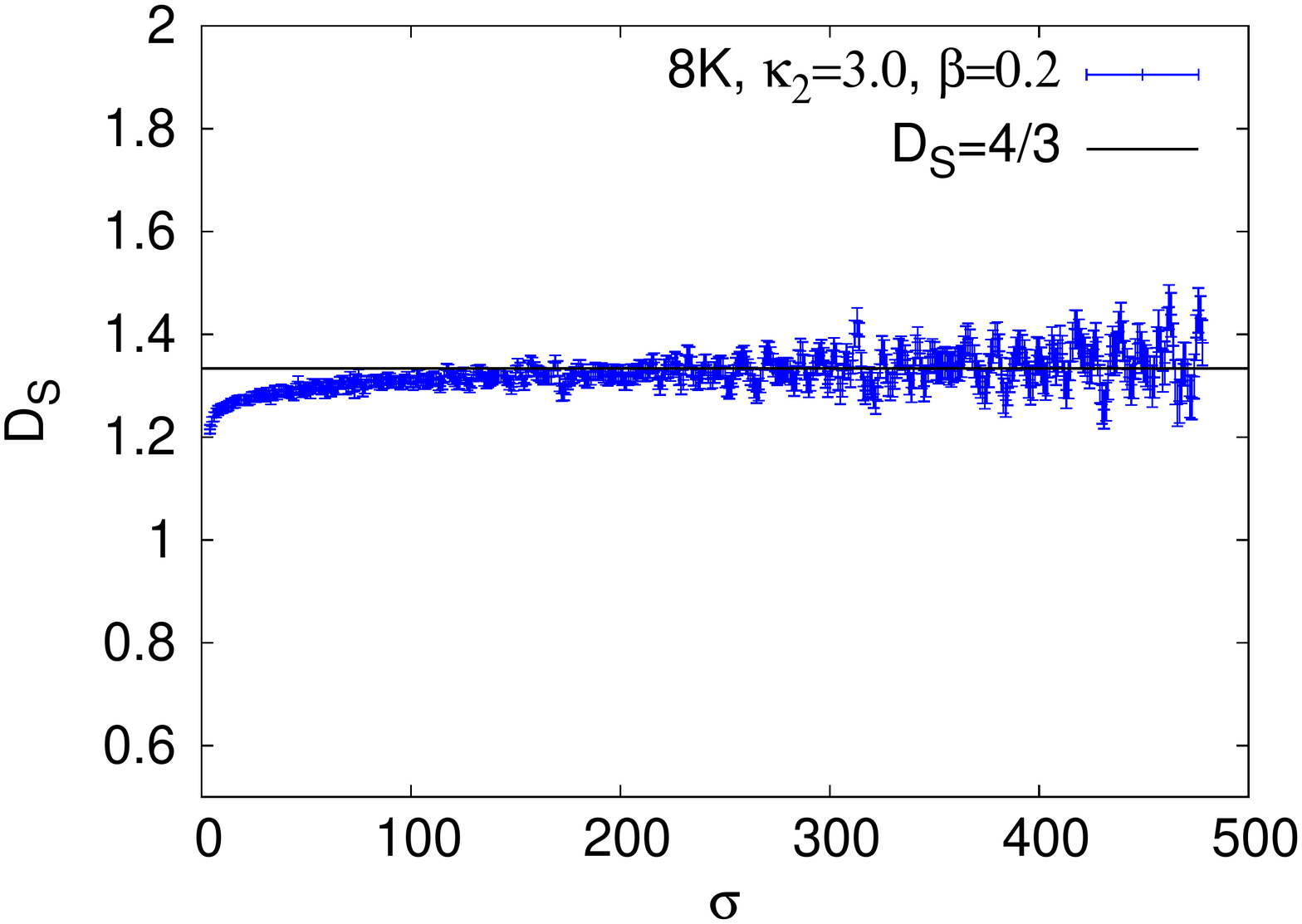}\label{fig:SpecBP8K}}\hspace{-1.5cm}
\subfloat[]{\includegraphics[width=0.55\textwidth,natwidth=610,natheight=642]{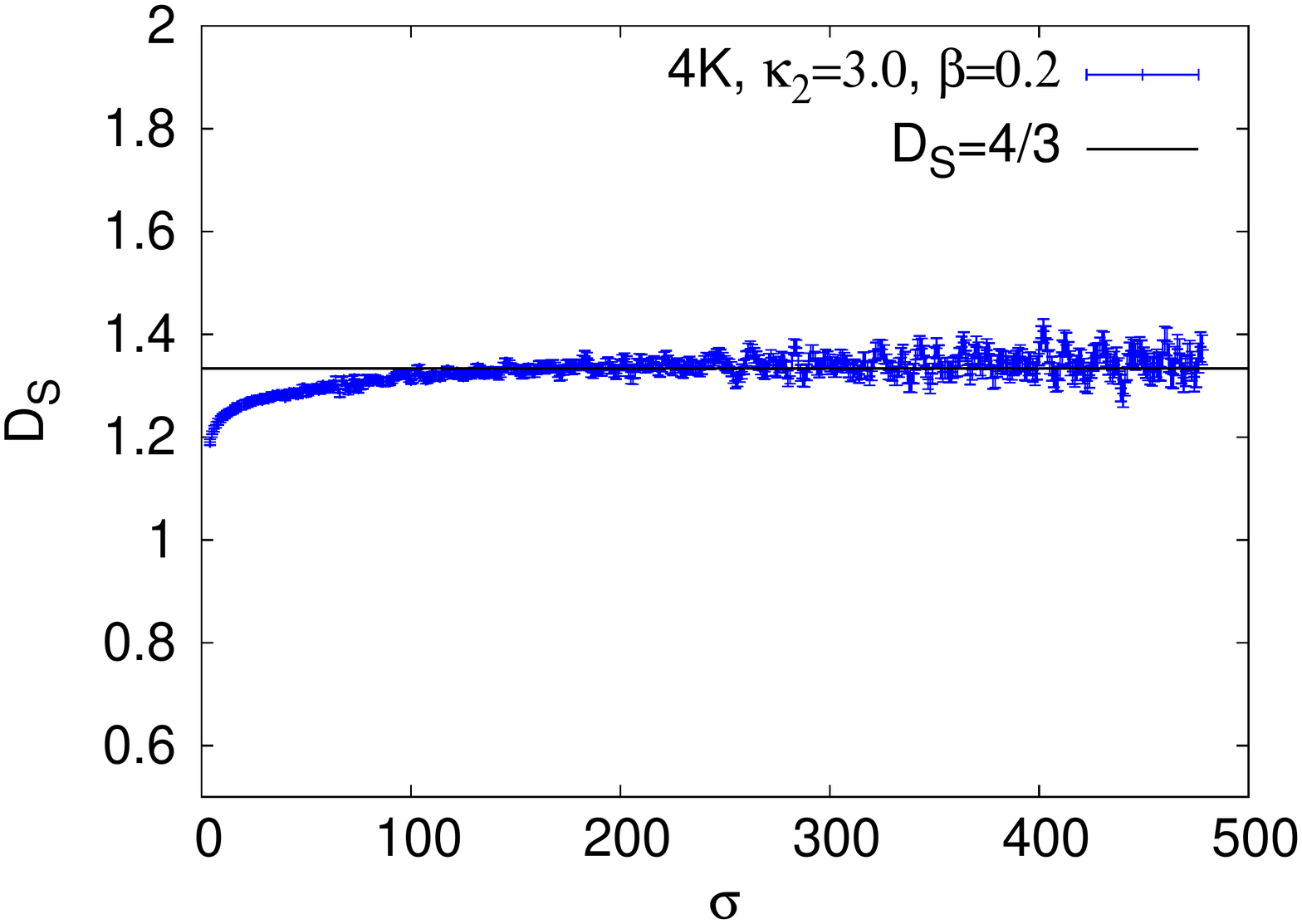}\label{fig:SpecBP4K}}\qquad
\subfloat[]{\includegraphics[width=0.55\textwidth,natwidth=610,natheight=642]{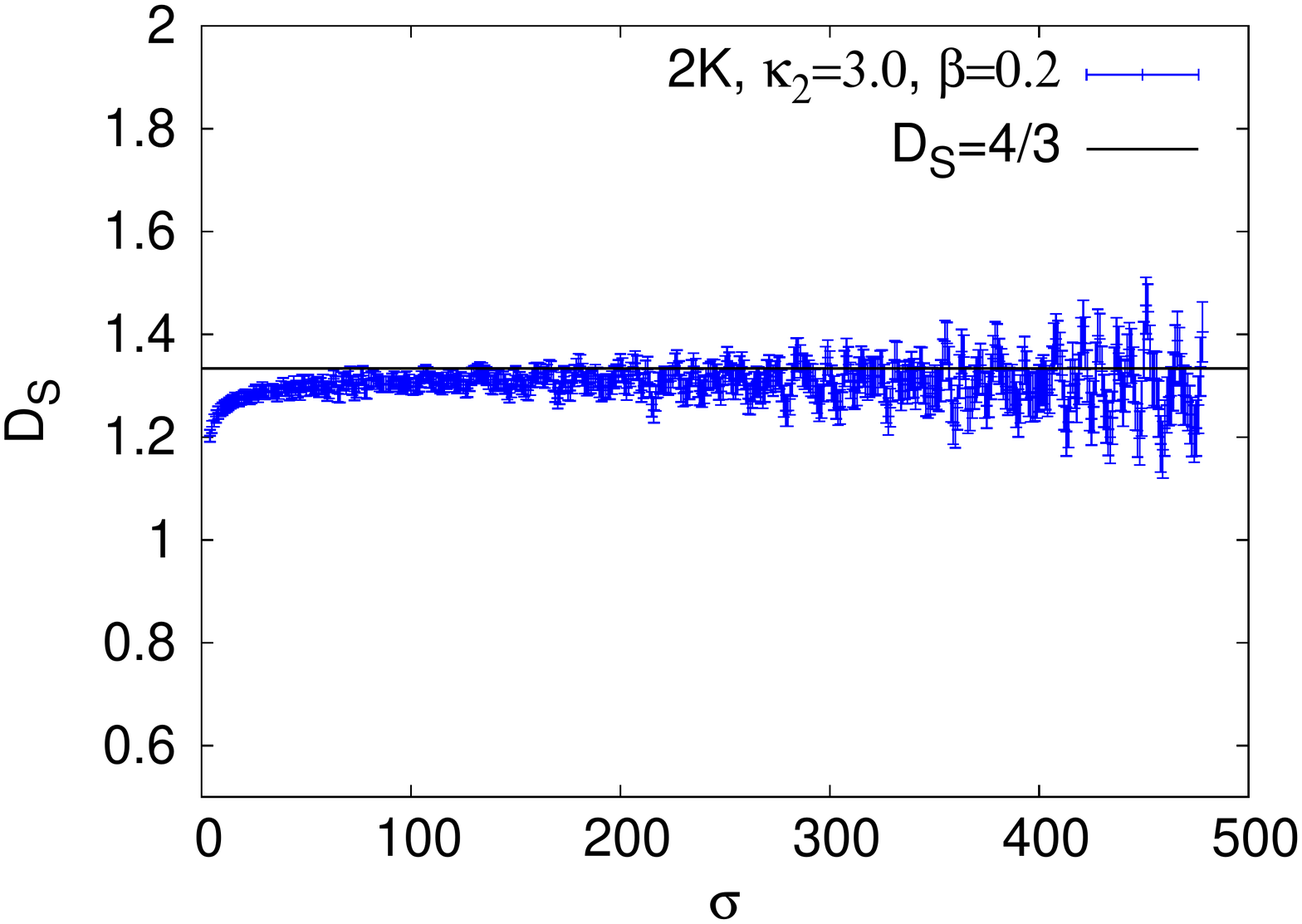}\label{fig:SpecBP2K}}\hspace{-1.5cm}
\subfloat[]{\includegraphics[width=0.55\textwidth,natwidth=610,natheight=642]{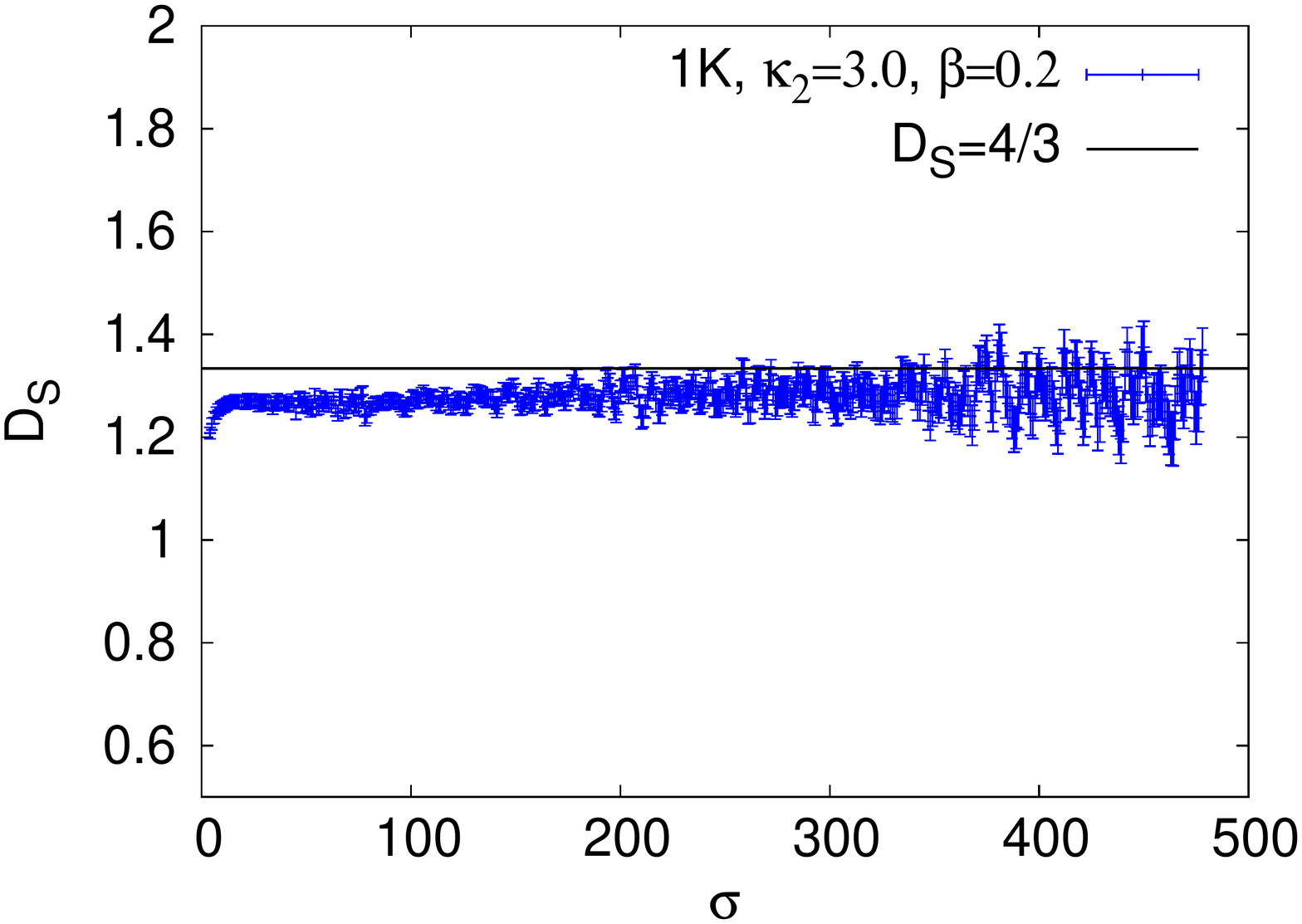}\label{fig:SpecBP1K}}             
\caption{\small The spectral dimension $D_{S}$ as a function of diffusion time $\sigma$ in the branched polymer phase ($\kappa_{2}=3.0$, $\beta=0.2$). The theoretical prediction of $D_{S}$=4/3 is shown in each plot for comparison. $D_{S}$ is calculated as a function of $\sigma$ for volumes (a) $N_{4}$=8000, (b) $N_{4}$=4000, (c) $N_{4}$=2000, and (d) $N_{4}$=1000.
}\label{fig:SpecBP}
\end{figure}

\begin{table}
\centering
\caption{A table of the volume $N_{4}$ of the ensemble and the spectral dimension $D_{S}$ determined on that ensemble.}
  \begin{tabular}{|c|c|}
    \hline
    Volume ($N_{4}$) & $D_{S}$ \\ \hline\hline
    8000 & $1.335(10)$ \\ \hline
    4000 & $1.342(7)$ \\ \hline
    2000 & $1.309(13)$ \\ \hline
    1000 & $1.285(12)$ \\
    \hline
  \end{tabular}
  \label{BPTable}
\end{table}

It is well-established in the literature \cite{deBakker:1996zx} that at $\beta=0$ the transition between the branched polymer phase and the collapsed phase (see Fig. \ref{PhaseDiagram}) is first-order for combinatorial triangulations, and Ref. \cite{Bilke:1998bn} provides evidence that this is also true for degenerate triangulations. Our simulations support this picture. We find a first-order phase transition at $\beta=0$ and similar discontinuous behavior along the line AB, though it is difficult to study the transition for negative values of $\beta$ below $\sim -0.5$ because of large finite-size effects. Evidence for a first-order transition is presented in Section \ref{firstorderphasetransition}.

The dashed line CD separating the collapsed phase and the crinkled region appears to be a much softer transition than the solid line AB. Our studies show continuous behavior at this line, which is indicative of an analytic cross-over. Within the crinkled region finite-size effects are much greater than they are for $\beta$=0, and this previously led us to think that this phase possessed a 4-dimensional extended geometry \cite{Laiho:2011ya,Coumbe:2012qr,Laiho2014}. However, simulations at larger volumes suggest that the crinkled region is a region within the collapsed phase with very large finite-size effects and long auto-correlation lengths. Thus for small lattice volumes the spectral dimension can appear to asymptote to 4 in the crinkled region. However, with increasing lattice volumes the spectral dimension increases beyond 4, just as we have seen here for the collapsed phase. Section \ref{crinkledphase} discusses our study of the crinkled region in more detail.              

\end{subsection}


\begin{subsection}{The Phase Transition Line A-B}\label{firstorderphasetransition}

One way to determine the order of a transition is to study the finite-size scaling of the susceptibility of a suitably chosen order parameter. A first-order phase transition in the infinite volume limit is discontinuous in the order parameter. A second-order transition in the infinite volume limit, however, is continuous in the order parameter but discontinuous in the first derivative \cite{Blundell:2008ca}. Since it is not possible to simulate in the infinite volume limit it can be difficult to distinguish numerically between first and second-order transitions. For both first and second-order transitions the peak height of the susceptibility of the order parameter diverges. However, the susceptibility diverges with a different scaling exponent, thus allowing one to distinguish between the two. For a cross-over the peak in the susceptibility does not diverge but remains constant as a function of volume.  

A suitable choice of order parameter for exploring the phase diagram is provided by the expectation value of the Regge curvature.

\begin{equation}\label{Rdef}
\langle R \rangle \equiv \frac{\langle\int d^{4}x\sqrt{g}R\rangle}{\langle \int d^{4}x\sqrt{g}\rangle}
\end{equation}

\noindent The average Regge curvature is straightforward to determine from the bulk variables $N_{2}$ and $N_{4}$, with

\begin{equation}\label{AvgR}
\left\langle R \right\rangle\approx\frac{1}{\rho} \left<\frac{N_{2}}{N_4}\right> - 1,
\end{equation}

\noindent where \begin{math}\rho=\frac{10\arccos\left(1/4\right)}{2\pi}\end{math} and \begin{math}N_{i}\end{math} is the total number of \emph{i}-simplices.  Although we follow Ref.~\cite{deBakker:1994zf} in using $N_2/N_4$ as our order parameter, note that Eq.~(\ref{AvgR}) is a strict equality only when $N_4$ is held fixed.  In our simulations $N_4$ is allowed to vary slightly, and so the identification in Eq.~(\ref{AvgR}) is not exact, though given the small variation in $N_4$, the difference between the exact and approximate expressions for $\left<R\right>$ is not significant.  Taking the derivative of the Regge curvature with respect to \begin{math}\kappa_{2}\end{math} gives the curvature susceptibility \begin{math}\chi_{R}\left(N_{2},N_{4}\right)\end{math} as a function of $N_{2}$ and $N_{4}$. The curvature susceptibility $\chi_{R}$ is given by
\begin{equation}\label{ChiR}
\chi_{R}(N_{2},N_{4})\approx\left[\left\langle \left(\frac{N_{2}}{N_4}\right)^{2}\right\rangle -\left\langle \frac{N_{2}}{N_4}\right\rangle ^{2}\right]N_4.
\end{equation}
As in Eq.~(\ref{AvgR}), Eq.~(\ref{ChiR}) is only exact when $N_4$ is strictly constant, but the variation of $N_4$ is small enough to make no qualitative difference in our study. We indicate the approximate nature of Eqs.~(\ref{AvgR}) and~(\ref{ChiR}) by adding a superscript ``ap'' to the axis labels of all plots that show data where this approximation was made. Most early studies of the phase diagram of EDT used $N_0$ as an order parameter (see e.g. \cite{Bialas:1996wu}), rather than $N_2/N_4$.  They studied the susceptibility defined by   
\begin{equation}  \chi_R(N_0,N_4) = \frac{1}{N_4}\left(\left<N_0^2\right>-\left<N_0\right>^2\right),
\end{equation}
which should agree with that of Eq.~(\ref{ChiR}) up to an overall factor for fixed $N_4$.  From the relation
\begin{equation} N_0 - \frac{1}{2}N_2 + N_4 = \chi,
\end{equation}
where $\chi$ is the Euler number and is equal to 2 for $S^4$, we find that the expected overall factor is 4. Figure~\ref{ChiComparison} shows good agreement between these two definitions of the susceptibility in the vicinity of the phase transition once this factor of 4 is taken into account.  Thus we conclude that $N_2/N_4$ serves as a good order parameter, and we use it throughout the remainder of this work.  

\begin{figure}[H]
  \centering
  \includegraphics[width=0.8\linewidth,natwidth=610,natheight=642]{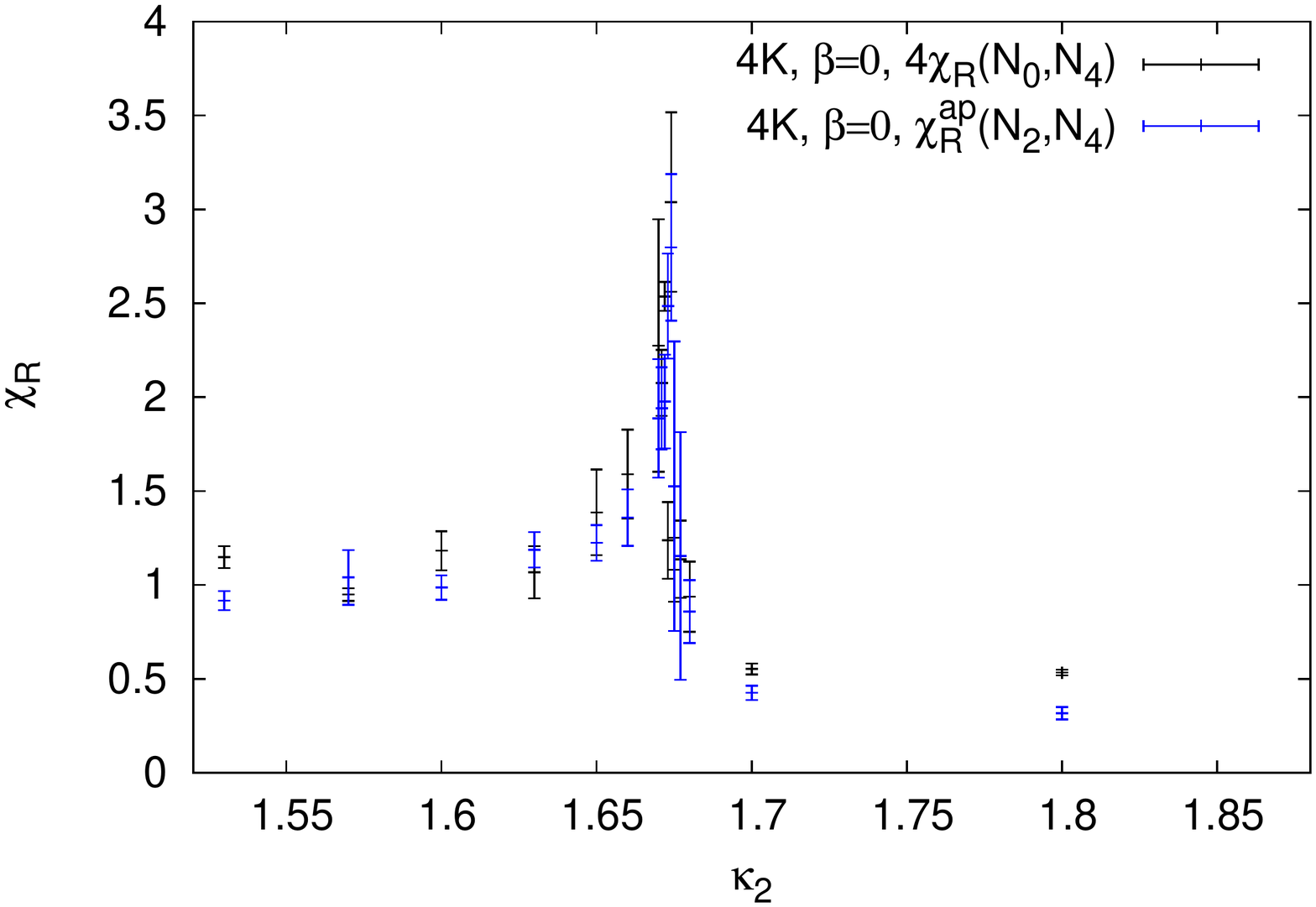}
  \caption{\small The curvature susceptibility $\chi_R$  as a function of $\kappa_{2}$ at 4K volume for two different definitions of $\chi_R$. The superscript ``ap'' indicates that the definition of susceptibility given by Eq.~(\ref{ChiR}) is approximate.}
\label{ChiComparison}
\end{figure}

In the rest of this section we study the curvature susceptibility $\chi_{R}$ with the aim of locating the transition line AB in the phase diagram. We study the order of the transition by examining the time history of the number of vertices in the Monte Carlo run. At a first-order transition the order parameter exhibits a number of discontinuous transitions between two meta-stable states as a function of Monte Carlo time. A histogram of the time history of $N_{0}$ from such a run is expected to give a double Gaussian distribution if the transition is first-order. Crucially, a double Gaussian distribution for just one lattice volume is not sufficient to confirm a first-order transition; one must observe such a distribution over at least two volumes and see the peak separation grow with volume, since only in the infinite volume limit does the transition become truly discontinuous. 

In the following subsections we consider a series of lines through the phase diagram with varying $\kappa_{2}$ and fixed $\beta$.


\begin{subsubsection}{$\bm{\beta}$=0}

We study the order parameter $R$ and its susceptibility $\chi_{R}$ as defined in Eqs. (\ref{AvgR}) and (\ref{ChiR}), respectively. Figure \ref{AvgR4K8KB0} shows the expectation value of the Regge curvature as a function of $\kappa_{2}$ for $\beta$ fixed at 0. We consider the case of $\beta=0$ first because finite-size effects increase as $\beta$ decreases and this provides our cleanest example of a first-order phase transition. It is also the case most studied previously in the literature. Figure \ref{AvgR4K8KB0} shows the average Regge curvature as a function of $\kappa_{2}$. As can be seen in Fig. \ref{AvgR4K8KB0}, for $\kappa_{2}\leq0.5$ the curvature is a slowly increasing function of $\kappa_{2}$ which then undergoes a rapid transition into another plateau region when $\kappa_{2}\geq2.0$. Since the curvature susceptibility $\chi_{R}$ is the first derivative of the Regge curvature with respect to $\kappa_{2}$ one observes a sharp spike at the transition, as shown in Fig. \ref{ChiR4K8KB0}.  Although the scaling of the peak height could be used in principle to determine the order of the phase transition, the lattice volume is still not large enough to perform this study reliably, though perhaps this could be achieved using the methods of Ref.~\cite{Smit:2013wua}, where EDT was recently reexamined.

\begin{figure}[H]
  \centering
  \includegraphics[width=0.8\linewidth,natwidth=610,natheight=642]{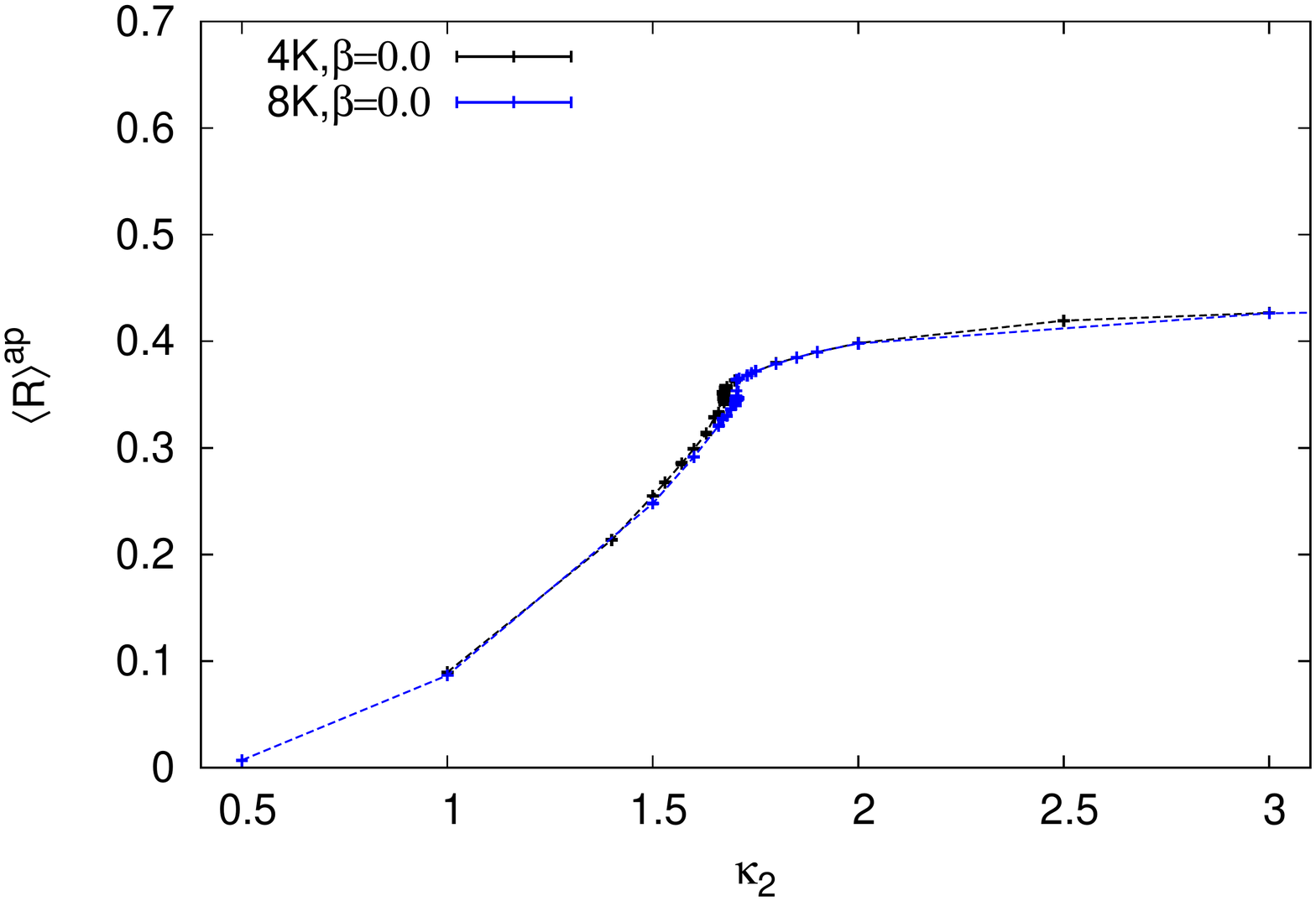}
  \caption{\small The average Regge curvature $\left\langle R \right\rangle^{ap}$ as a function of $\kappa_{2}$ for 4K and 8K volumes at $\beta=0$.}
\label{AvgR4K8KB0}
\end{figure}

\begin{figure}[H]
  \centering
  \includegraphics[width=0.8\linewidth,natwidth=610,natheight=642]{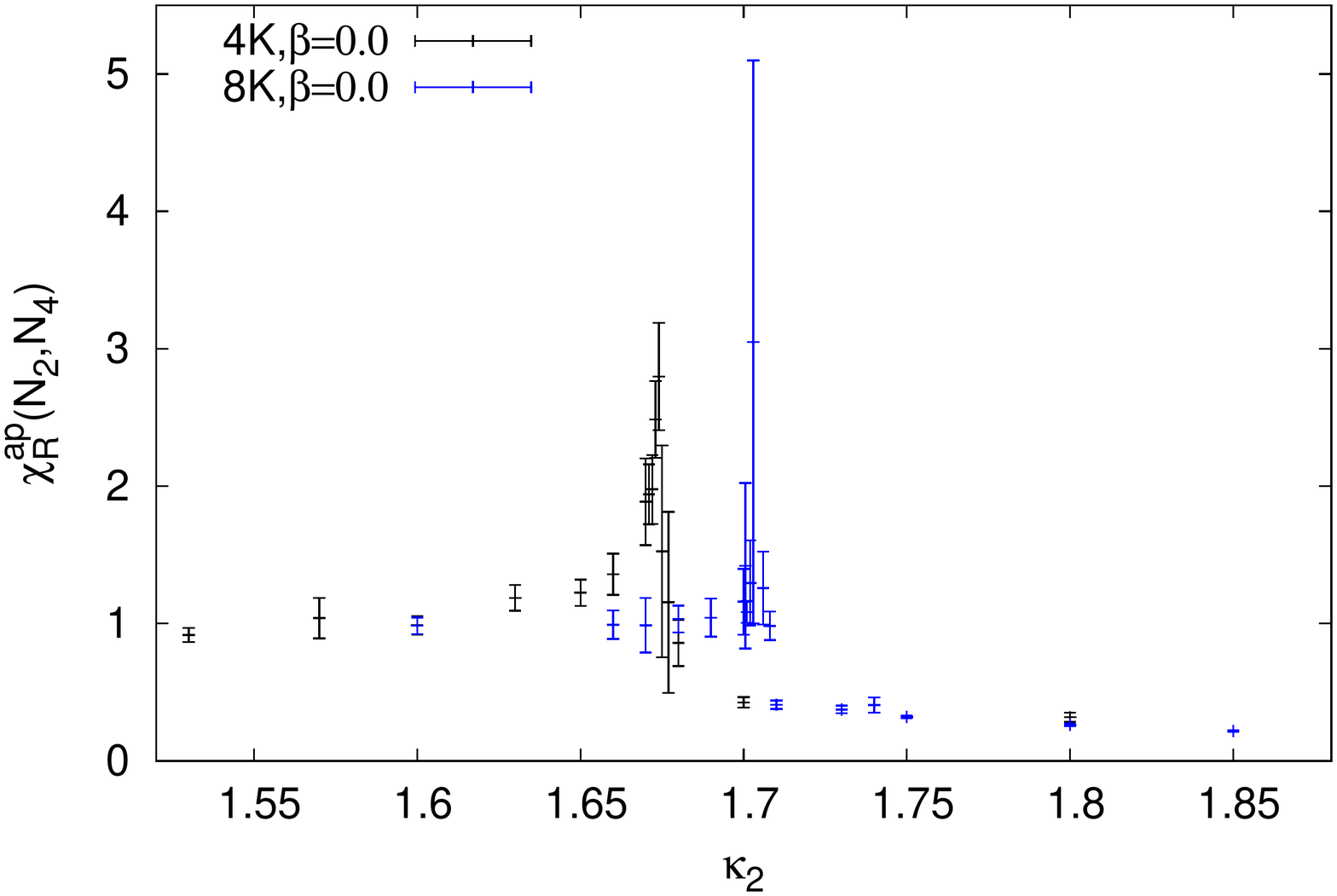}
  \caption{\small The curvature susceptibility $\chi_{R}^{ap}$ as a function of $\kappa_{2}$ for 4K and 8K volumes at $\beta=0$.}
\label{ChiR4K8KB0}
\end{figure}

The Monte Carlo time history of $N_{0}$ and the corresponding histograms at the transition points are plotted in Figs~\ref{MCtime4KB0}, \ref{Hist05KB0}, and \ref{Hist4K8KB0}. The (pseudo)critical $\kappa_{2}$ values at which the transition occurs are volume dependent, as expected from a first-order transition\interfootnotelinepenalty=10000\footnote{\scriptsize As an example consider a temperature-driven phase transition in a $d-$dimensional Ising model with volume $V=L^{d}$ \cite{NewmanBook}. From this one finds a power-law behavior for the transition point $p_{crit}$, which is given by $|p_{crit}\left(\infty\right)-p_{crit}\left(V\right)|\propto V^{-1/\tilde\nu}$, with $\tilde\nu=1$ for a first-order transition \cite{Ambjorn:2011cg, MeyerOrtmanns:1996ea}.}, with $\kappa^{c}_{2}(0.5K)=1.47$, $\kappa^{c}_{2}(4K)=1.669$ and $\kappa^{c}_{2}(8K)=1.7068$. Figure \ref{MCtime4KB0} shows evidence for discontinuous fluctuations between the two metastable states for the 4K and 8K ensembles; this behavior is characteristic of a first-order phase transition.  It is evident in these plots that there are multiple transitions between the metastable states for both the 4K and 8K ensembles, but not for our smaller 0.5K run. To improve statistics on the 8K ensemble we have run multiple parallel streams in addition to the one shown in Fig.~\ref{MCtime4KB0}.   The histogram in Fig. \ref{Hist4K8KB0} includes all of the 8K parallel streams, where we have started measuring after sweep 10,000 on each stream in order to ensure that each is thermalized.  Discontinuous fluctuations are not observed for the much smaller 0.5K ensemble, as can be seen in Fig.~\ref{MCtime4KB0}.  Figure~\ref{Hist05KB0} shows the histogram of $N_{0}$ (divided by the volume $N_{4}$) for the 0.5K ensemble.  No double peak structure is observed for this ensemble. Figure \ref{Hist4K8KB0} shows the histogram for the two larger volumes, where a clear double Gaussian peak is observed that becomes more pronounced as the volume increases.  The distance between the peaks grows approximately linearly with volume for a first-order transition, so the peak separation for $N_{0}/N_{4}$ should be a constant for a first-order transition. Figure \ref{Hist4K8KB0} is compatible with this picture, thus providing strong evidence that the transition between the collapsed and branched polymer phases is first-order.  The much smaller volume does not show any evidence of a double peak in the histogram, but this only strengthens the argument that the transition is first-order, since the separation in the peaks is expected to become more pronounced as the volume is increased for a first-order transition; this is what is seen in our data.  This result is already well-established in the literature for combinatorial triangulations \cite{deBakker:1996zx}. The conclusion that the transition is first-order for degenerate triangulations in Ref. \cite{Thorleifsson:1998nb} was based on a double Gaussian in the time history of $N_{0}$ at a single volume only. Our result at a second larger volume bolsters this conclusion. Since combinatorial and degenerate triangulations agree on this result it lends support to the assumption that the two classes of triangulations have the same salient features, and to the assumption that if a suitable modification could be found that gave rise to a second-order transition, they would be in the same universality class.

\begin{figure}[H]
  \centering
  \subfloat{\includegraphics[width=0.85\linewidth,natwidth=610,natheight=642]{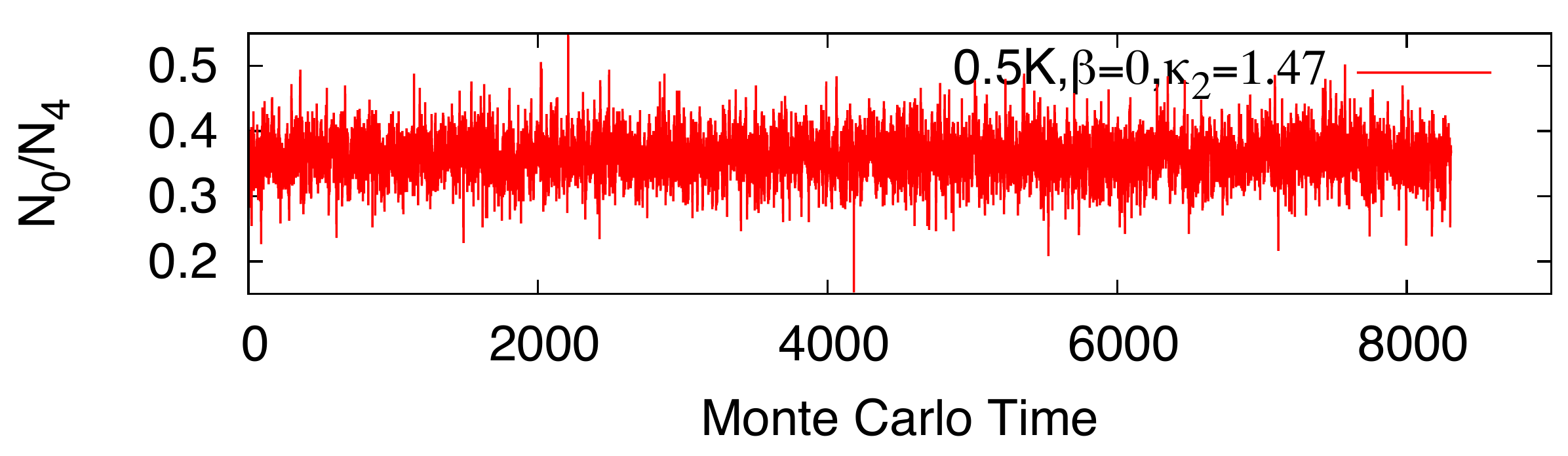}\label{MCtime05KB0}}\\
  \subfloat{\includegraphics[width=0.85\linewidth,natwidth=610,natheight=642]{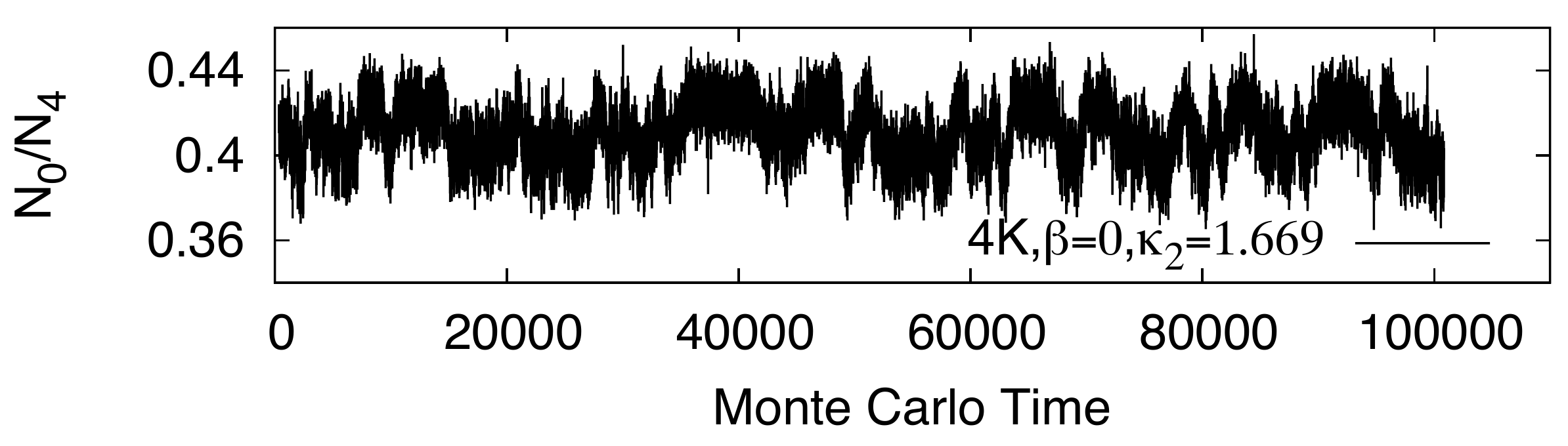}\label{MCtime4KB0}}\\
  \subfloat{\includegraphics[width=0.85\linewidth,natwidth=610,natheight=642]{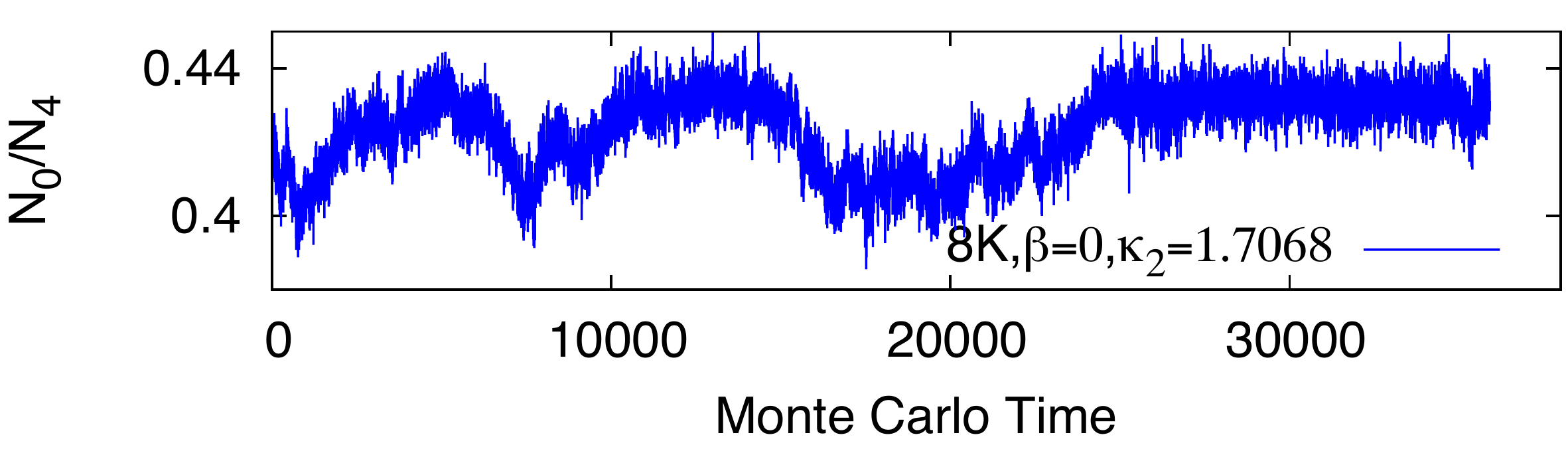}\label{MCtime8KB0}}
  \caption{\small The Monte Carlo time history (in units of $10^{8}$ attempted moves) of $N_{0}/N_{4}$ for 0.5K, 4K and 8K volumes at $\beta=0$. The pseudo-critical $\kappa_{2}$ value for the 0.5K ensemble (top figure) at $\beta=0$ is $\kappa_{2}=1.47$, for the 4K ensemble it is $\kappa_{2}=1.669$, and for the 8K ensemble (bottom figure) $\kappa_{2}=1.7068$.}
  \label{MCtime4KB0}
\end{figure}

\begin{figure}[H]
  \centering
  \includegraphics[width=0.8\linewidth,natwidth=610,natheight=642]{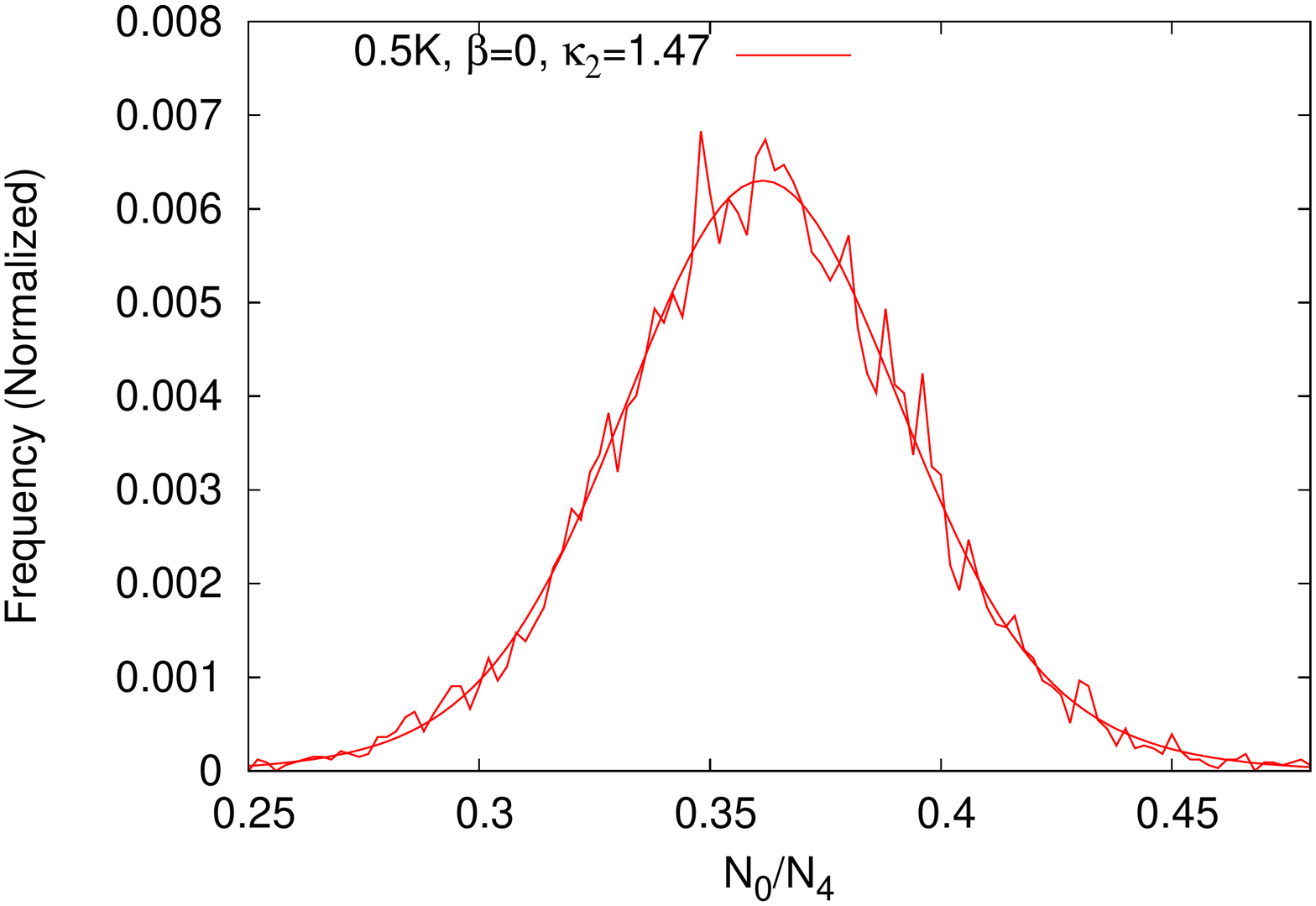}
  \caption{\small The histogram of $N_{0}/N_{4}$ including a double Gaussian fit for the 0.5K volume at $\beta=0$.  The frequency is normalized by dividing by the total number of entries in the histogram. }
\label{Hist05KB0}
\end{figure}

\begin{figure}[H]
  \centering
  \includegraphics[width=0.8\linewidth,natwidth=610,natheight=642]{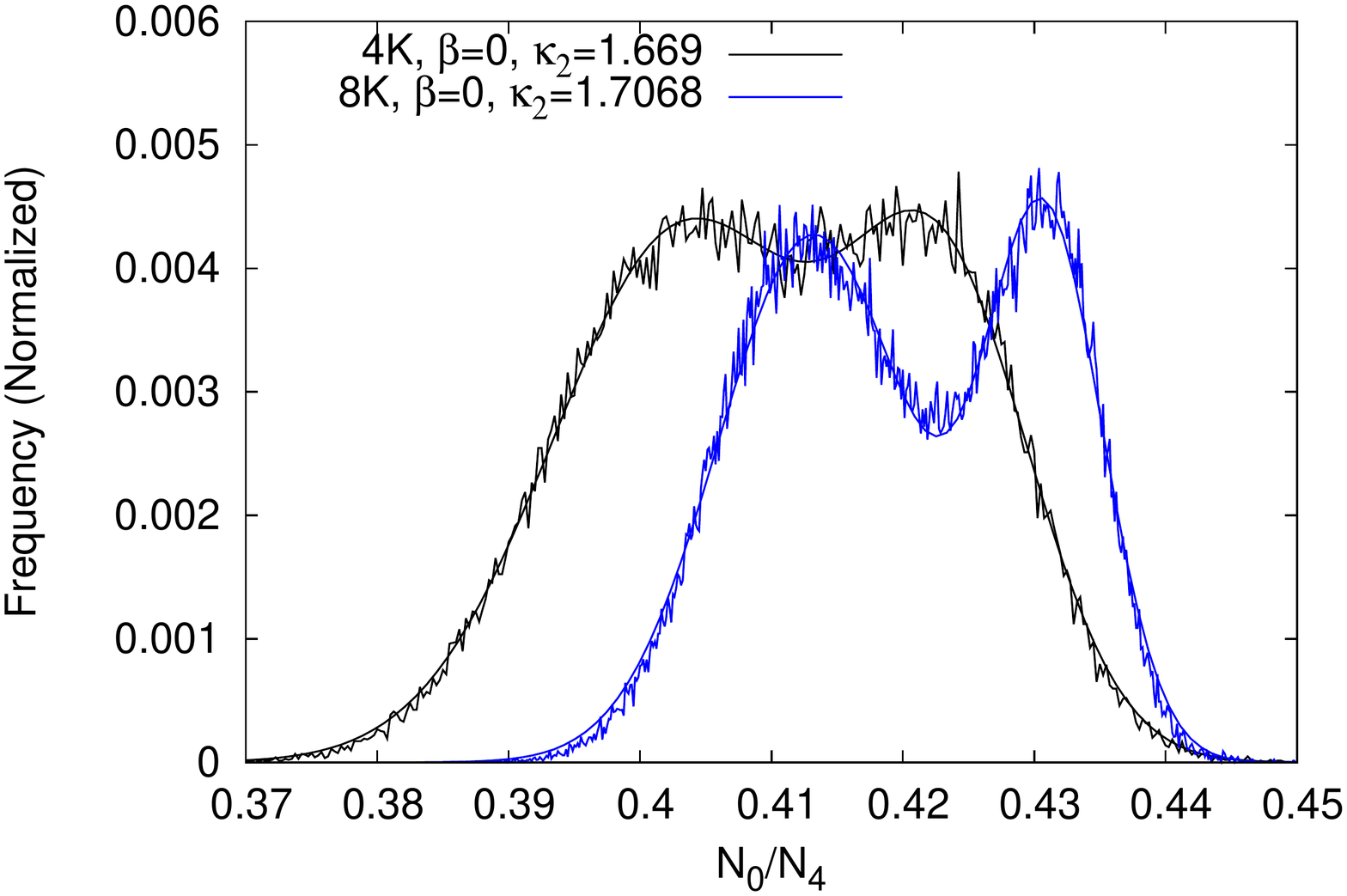}
  \caption{\small The histogram of $N_{0}/N_{4}$ including double Gaussian fits for the 4K and 8K volumes at $\beta=0$.  The frequency is normalized such that the peak heights of the two curves are approximately equal.}
\label{Hist4K8KB0}
\end{figure}


\end{subsubsection}


\begin{subsubsection}{$\bm{\beta}$=-0.6}

As $\beta$ becomes more and more negative the metropolis acceptance decreases. Also, the runs take longer to thermalize, especially in the transition region, where we experience critical slowing-down. When combined with increased finite-size effects these difficulties make it hard to determine the order of the transition and the location of the critical line. The presence of a peak in the susceptibility where the autocorrelation times become extremely long suggests that a genuine phase transition persists, at least down to $\beta\approx -0.6$.  



Figure~\ref{AvgR4K8KB06} shows the average Regge curvature as a function of $\kappa_2$ along the $\beta=-0.6$ line.  One can see that the slope in $\langle R \rangle$ is not as steep as for $\beta=0$.  For $\beta=-0.6$ the large finite-size effects and long auto-correlation lengths make it difficult to locate the position of the critical $\kappa_{2}$ value at which the transition occurs. We have only one volume $N_{4}=4000$, because larger lattices have still longer autocorrelation times and take too long to thermalize with available computing resources and algorithms.  The transition point at 4K is in the range $\kappa_{2}=2.4$-$2.5$, which we infer from Figs.~\ref{ChiR4K8KB06} and \ref{ChiR4KB0B02B04B06}.  Figure~\ref{ChiR4K8KB06} shows a small peak that deviates from the background around $\kappa_2=$2.4-2.5.  Although the significance of this deviation is not very great, perhaps 2-3$\sigma$, this point has a dramatically longer autocorrelation time than its neighbors, another sign of a phase transition.  This location of the transition for $\beta=-0.6$ is also around the region that one would infer from the trend in $\beta$ from Figure~\ref{ChiR4KB0B02B04B06}.  For more negative $\beta$ values, no sign of the transition is seen for the values of $\kappa_2$ and $\beta$ that we have explored so far.


\begin{figure}[H] 
  \centering
  \includegraphics[width=0.8\linewidth,natwidth=610,natheight=642]{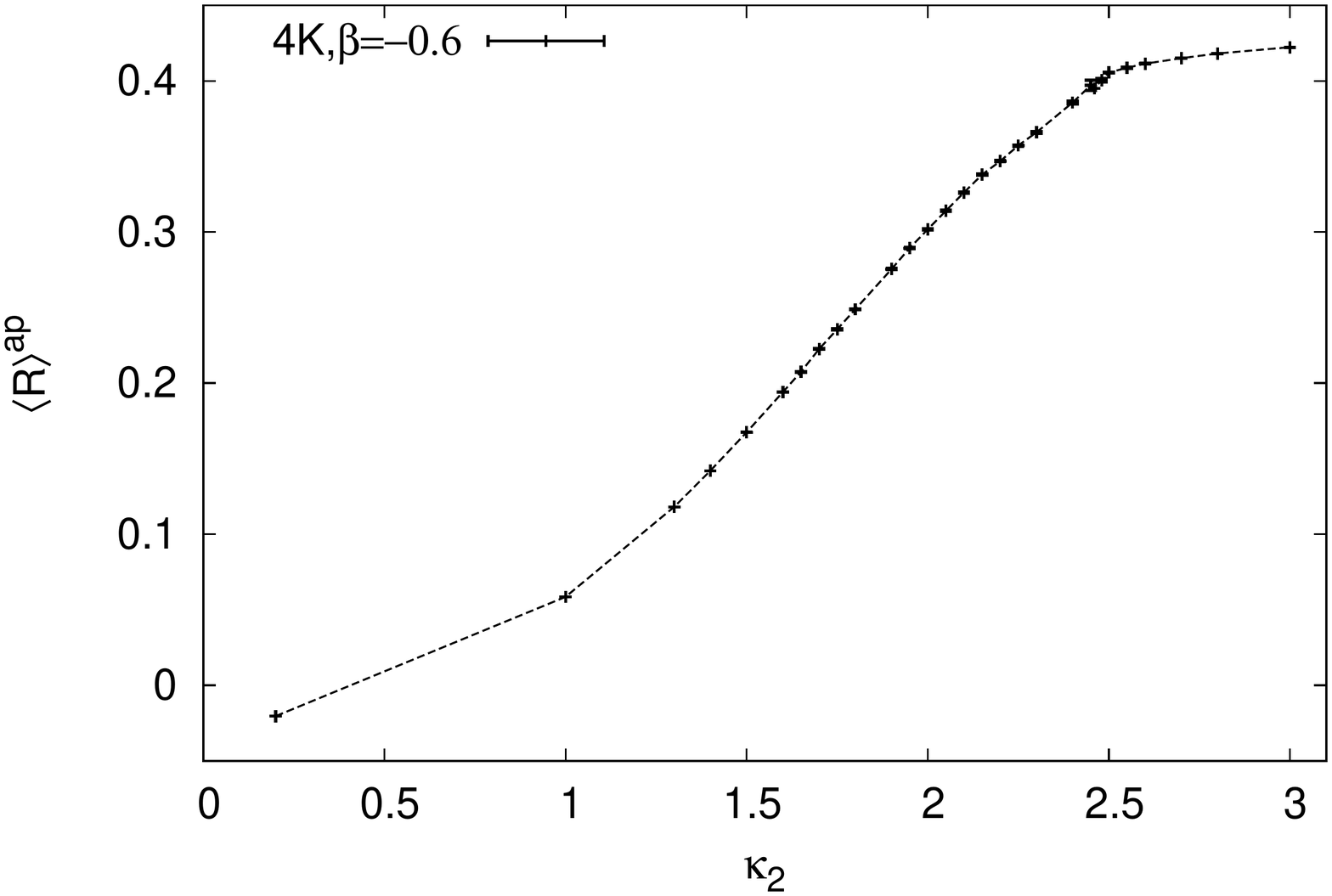}
  \caption{\small The average Regge curvature as a function of $\kappa_{2}$ for the 4K volume at $\beta=-0.6$.}
\label{AvgR4K8KB06}
\end{figure} 

\begin{figure}[H] 
  \centering
  \includegraphics[width=0.8\linewidth,natwidth=610,natheight=642]{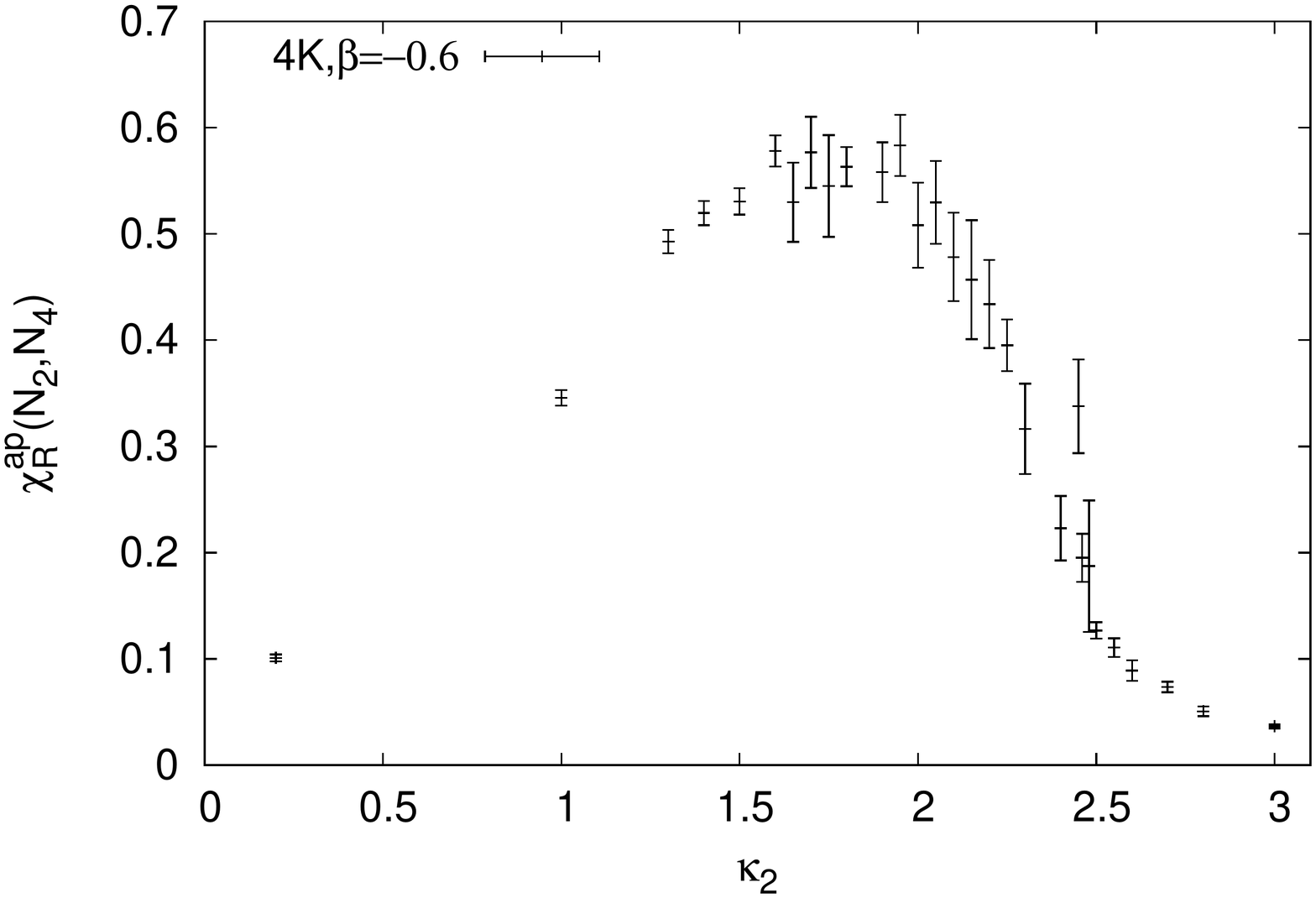}
  \caption{\small The curvature susceptibility $\chi_{R}^{ap}$ as a function of $\kappa_{2}$ for the 4K volume at $\beta=-0.6$.}
\label{ChiR4K8KB06}
\end{figure} 

\end{subsubsection}


\begin{subsubsection}{The Location of the Transition Line A-B in the Phase Diagram}

In this section we combine all the constraints on the location of the $\kappa^{c}_{2}$ values at which the transition occurs. This allows us to draw a more precise phase diagram, rather than a mere schematic. 

\begin{figure}[H] 
  \centering
  \includegraphics[width=0.8\linewidth,natwidth=610,natheight=642]{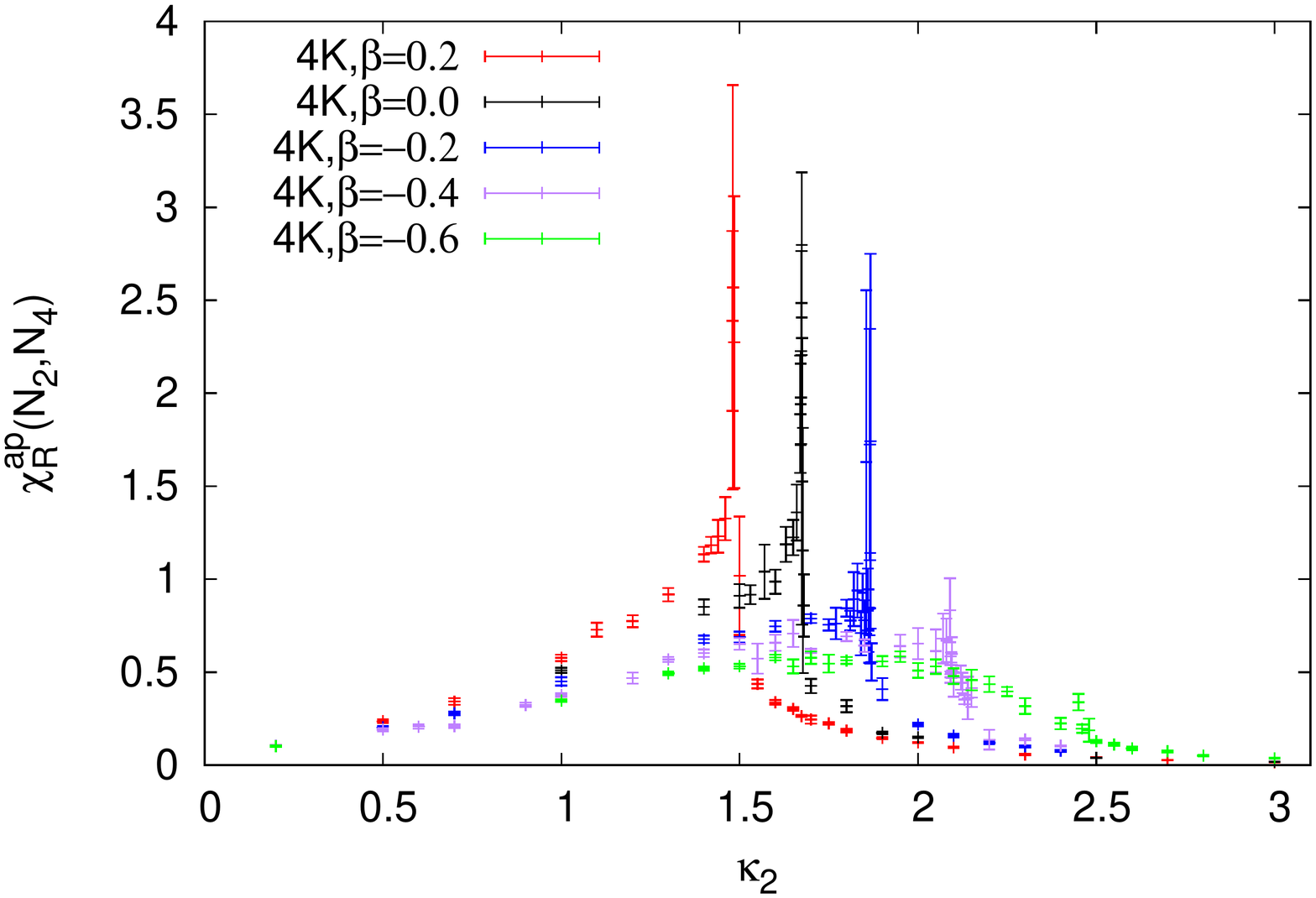}
  \caption{\small The curvature susceptibility $\chi_{R}^{ap}$ as a function of $\kappa_{2}$ for the 4K volume at $\beta$=0.2, 0, -0.2, -0.4, and -0.6.}
\label{ChiR4KB0B02B04B06}
\end{figure}       

\noindent Figure \ref{ChiR4KB0B02B04B06} shows that as we decrease $\beta$ the $\kappa^{c}_{2}$ values shift to the right. This behavior is shown schematically by line AB in Fig. \ref{PhaseDiagram}. Figure \ref{PDConstraints} shows the actual measurements of $\kappa^{c}_{2}$ and their associated errors at two different volumes. If a transition is present for more negative values of $\beta$, it is not visible in the data, and this leads to the one-sided bounds in Fig. \ref{PDConstraints}.

\begin{figure}[H] 
  \centering
  \includegraphics[width=0.8\linewidth,natwidth=610,natheight=642]{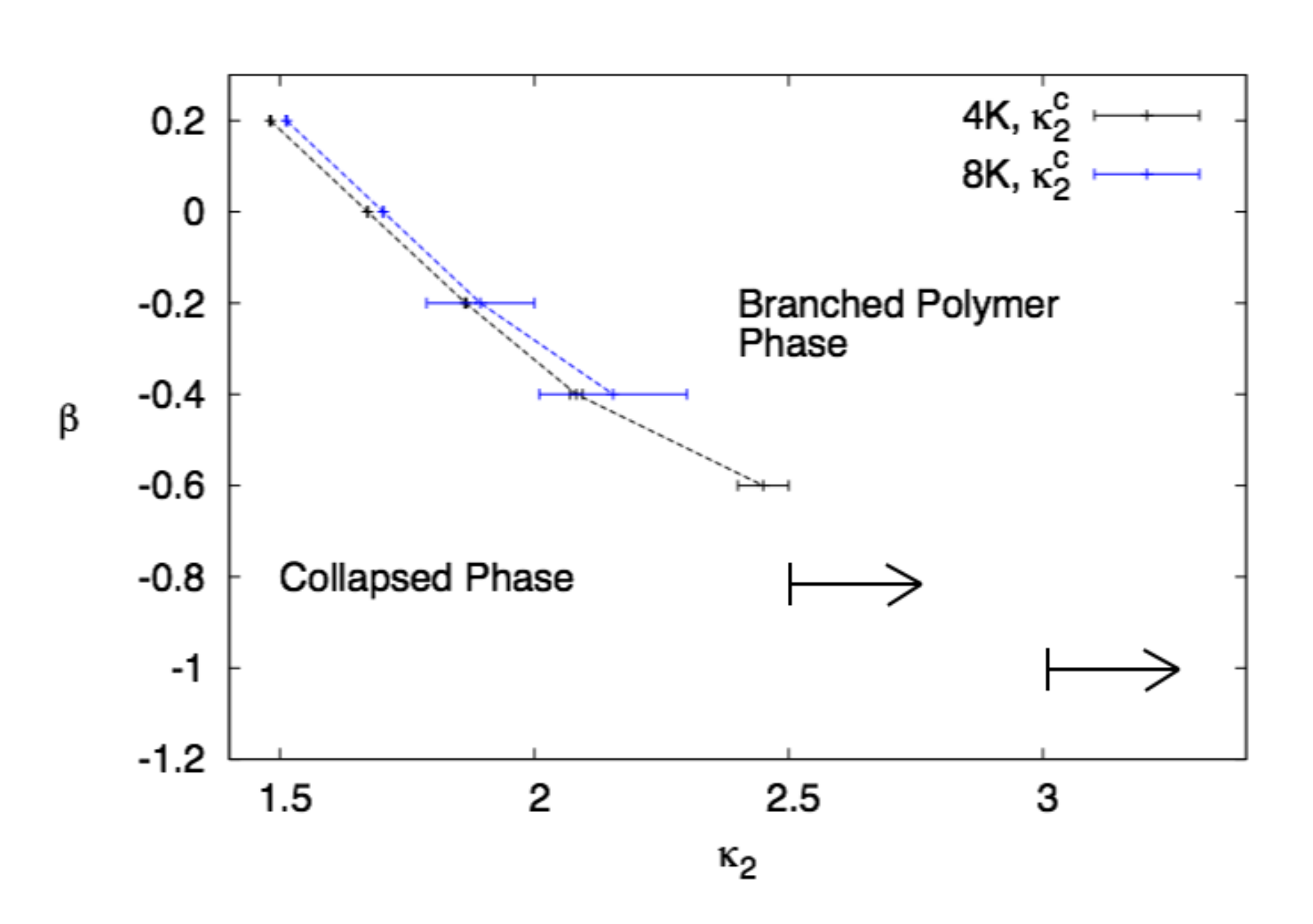}
  \caption{\small Constraints on the location of the first-order transition in the phase diagram of EDT with a non-trivial measure term based on the critical $\kappa_{2}$ values for the 4K and 8K lattice volumes.}
\label{PDConstraints}
\end{figure}       

Although it is difficult to determine the order of the transition for negative values of $\beta$,
mean field arguments suggest that the transition line remains first-order for negative values of $\beta$ in both three and four-dimensional EDT \cite{Bilke:1998vj}. Finite-size scaling, and renormalization group arguments based on the method of node decimation for 3-dimensional Euclidean dynamical triangulations with a nontrivial measure support this picture \cite{Warner:1998qa,Renken:1997na}. The current work is consistent with this possibility. 
 
\end{subsubsection}
\end{subsection}

\begin{subsection}{The Cross-over Line C-D}

The behavior of the curvature susceptibility as a function of $\kappa_{2}$ for $\beta=-1$ is consistent with an analytic cross-over. As expected at a cross-over, neither the height nor the width of the peak in the curvature susceptibility changes as a function of volume within errors. This is shown in Fig. \ref{ChiR4K8KB1}. If the phase diagram has a true transition somewhere along the $\beta=-1$ line it is most-likely at a larger value of $\kappa_{2}$, and to see it would require larger volumes than we can simulate with current methods.

We contrast the cross-over like behavior at $\beta=-1$ with the behavior at $\beta=0$ for very small volumes.  The results at $\beta=0$ for volumes of 200 and 500 4-simplices are shown in Figure~\ref{ChiR02K05K4KB0}, along with our results at volumes of 4000 and 8000 4-simplices in order to study how the expected behavior at the phase transition breaks down when the volume is sufficiently small.  Figure~\ref{ChiR02K05K4KB0} shows that the first-order transition observed in the curvature susceptibility at $\beta=0$ is difficult to detect when the lattice volume is very small. The peak becomes rather broad and smooth, resembling a cross-over rather than a true phase transition.  However, the susceptibility has a strong volume dependence, as can be seen in Fig. \ref{ChiR02K05K4KB0} by comparing the runs with 200 and 500 simplices. This is in contrast to the data at $\beta=-1$, which show essentially no finite-volume dependence, suggesting that the transition between the collapsed phase and the crinkled region is not a true phase transition.

\begin{figure}[H] 
  \centering
  \includegraphics[width=0.8\linewidth,natwidth=610,natheight=642]{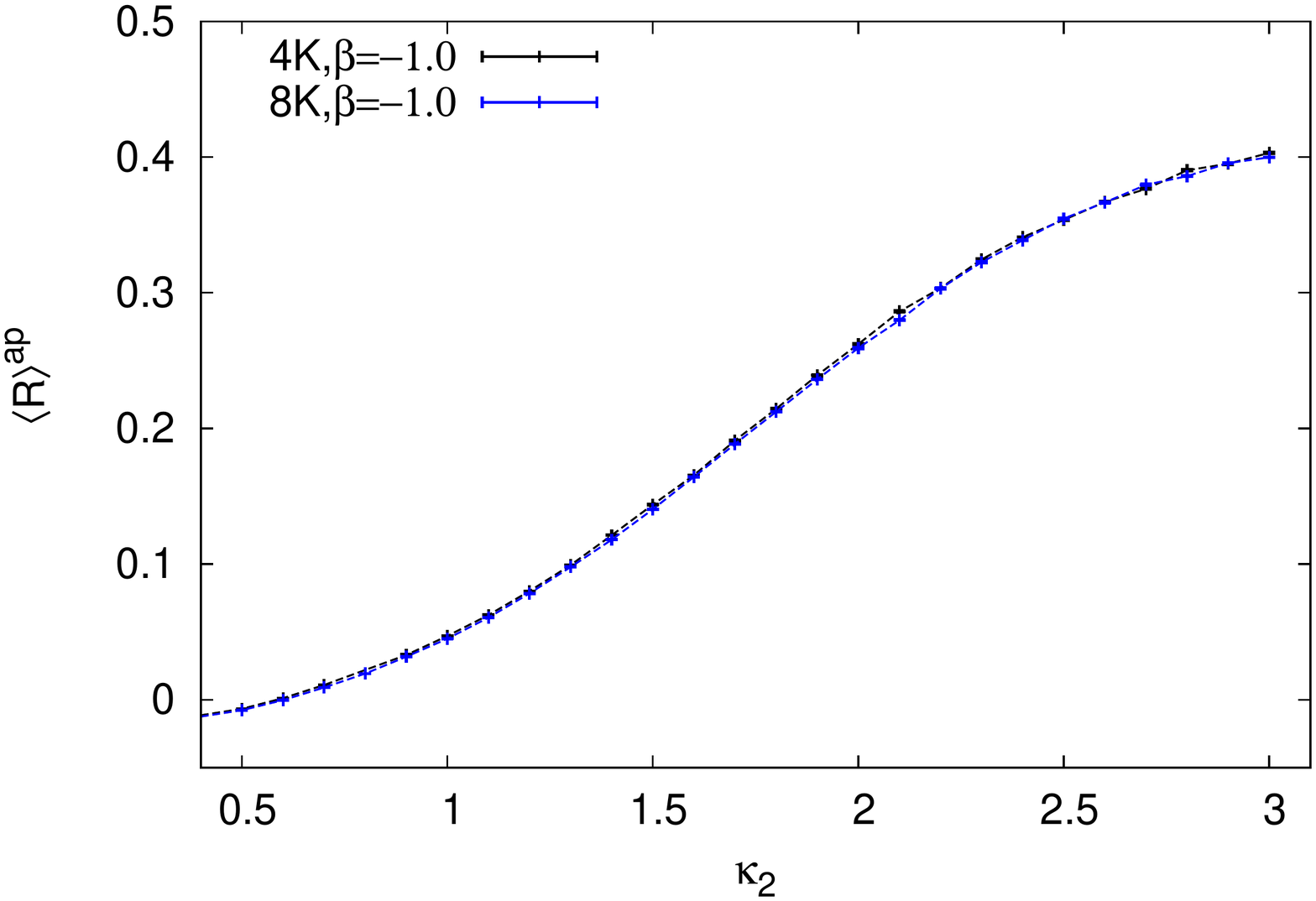}
  \caption{\small The Average Regge curvature as a function of $\kappa_{2}$ for the 4K and 8K volumes at $\beta=-1$.}
\label{AvgR4K8KB1}
\end{figure} 

\begin{figure}[H] 
  \centering
  \includegraphics[width=0.8\linewidth,natwidth=610,natheight=642]{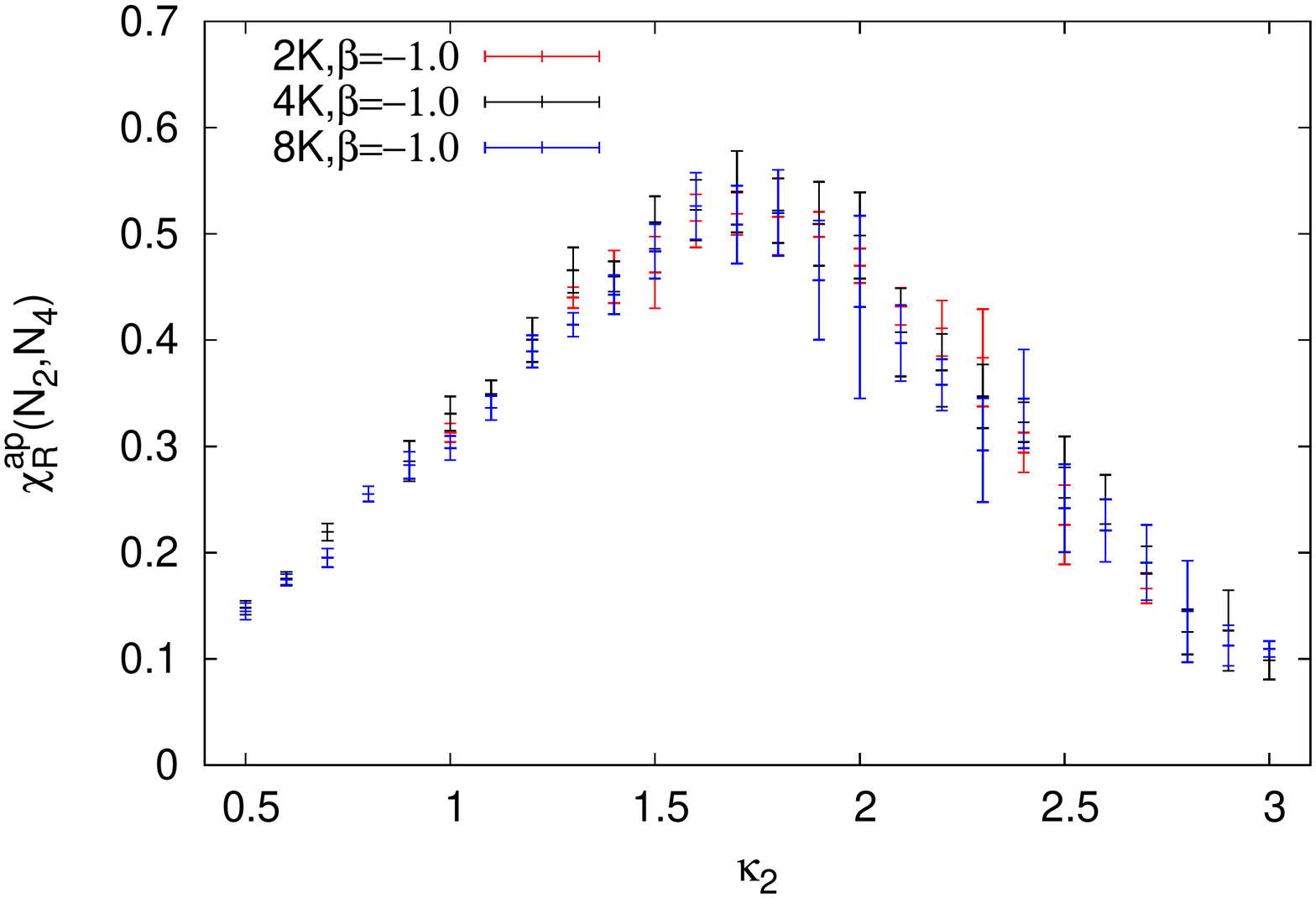}
  \caption{\small The curvature susceptibility $\chi_{R}^{ap}$ as a function of $\kappa_{2}$ for 2K, 4K, and 8K volumes at $\beta=-1$.}
\label{ChiR4K8KB1}
\end{figure}

 \begin{figure}[H] 
  \centering
  \includegraphics[width=0.8\linewidth,natwidth=610,natheight=642]{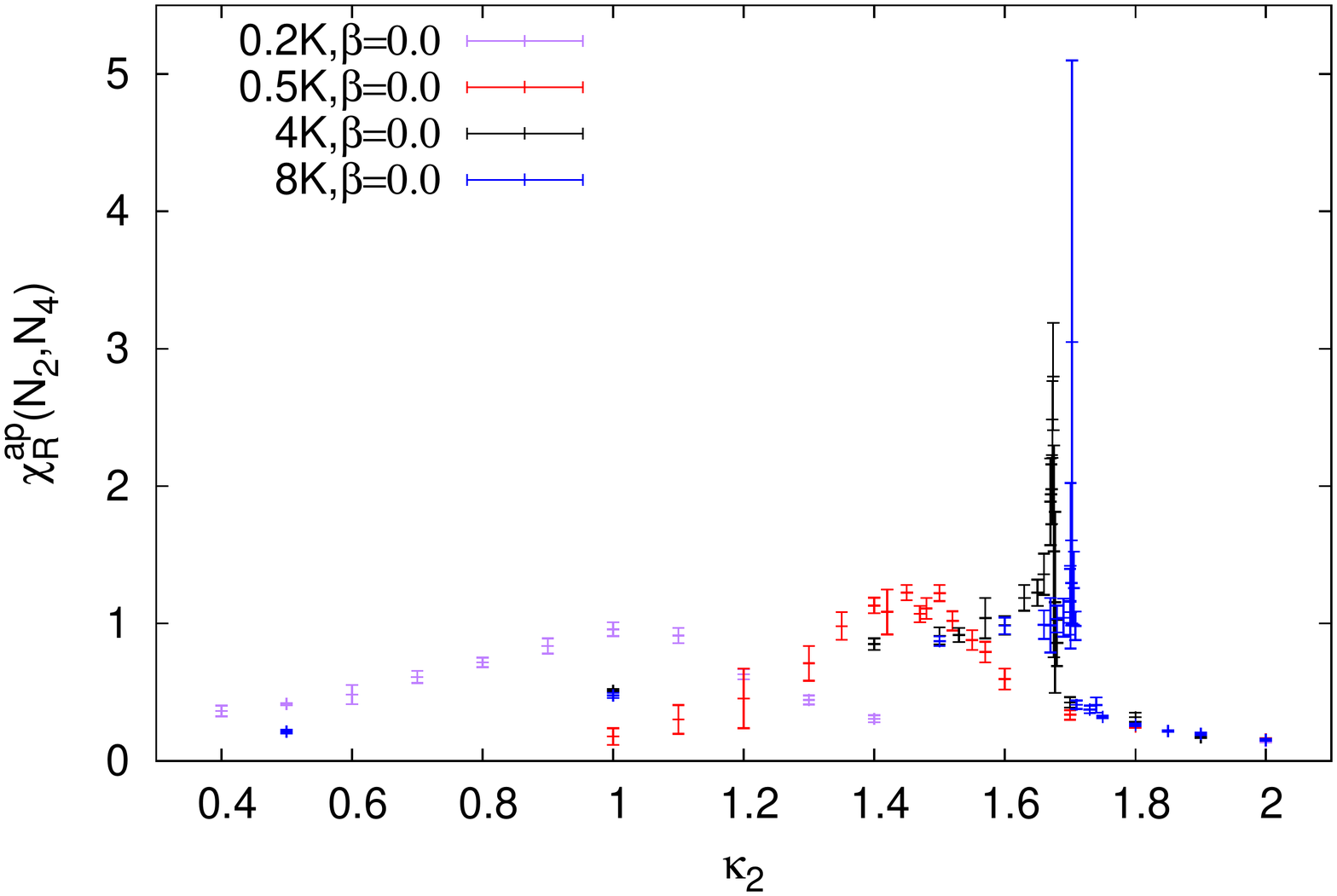}
  \caption{\small The curvature susceptibility $\chi_{R}^{ap}$ as a function of $\kappa_{2}$ for 0.2K, 0.5K, 4K, and 8K volumes at $\beta=0$. This plot shows that a sharp peak in the susceptibility only becomes evident for sufficiently large lattice volumes.}
\label{ChiR02K05K4KB0}
\end{figure}

\end{subsection}

\end{section}
\clearpage

\begin{section}{The Crinkled Region}\label{crinkledphase}

\begin{subsection}{Overview}

One of our main aims in studying the original formulation of EDT including a non-trivial measure term is to investigate the possibility that there exists a phase similar to the 4-dimensional semi-classical de Sitter phase found in CDT \cite{Ambjorn:1998xu,Ambjorn:2008wc}. The crinkled region in the phase diagram of EDT had not received much scrutiny in earlier studies where it was first identified \cite{Bilke:1998vj,Thorleifsson:1998qi}.  We reported in Ref.~\cite{Laiho:2011ya} that in the crinkled region, the spectral dimension has qualitatively similar behavior to that of CDT, including a running with distance scale and a short distance value that is compatible with $\sim 3/2$ \cite{Laiho:2011ya}.  We argued that the short-distance value for the spectral dimension of 3/2 might resolve the tension between asymptotic safety and black hole entropy scaling.  Given these intriguing results, it is important to take a closer look at the crinkled region of the EDT formulation.  Unfortunately, we find that with larger volumes we are not able to recover the desirable properties of the semiclassical phase of CDT within the crinkled region of EDT.  Our main evidence for this is our study of the Hausdorff dimension using finite-size scaling and our results for the spectral dimension on larger lattices than were available for our initial study.  In this section we present these new results for the Hausdorff and spectral dimensions within the crinkled region. The results support the conclusion that the crinkled region behaves like the collapsed phase but with larger finite-size effects, and there is no evidence that the crinkled region behaves like the 4-dimensional extended phase seen in CDT simulations.  This conclusion is consistent with the recent study of Ref.~\cite{Ambjorn:2013eha} for EDT with combinatorial triangulations.  

\end{subsection}

\begin{subsection}{The Hausdorff Dimension}\label{threevolumecorrelator}

We study the Hausdorff dimension in the crinkled region of our EDT simulations using the volume-volume correlator defined in Eq. (\ref{cvol}). Figure \ref{VolCorr4K8K12KB1} is a plot of the rescaled correlator $c_{N_{4}}(x)$ as a function of the rescaled linear dimension $x$, for six different lattice volumes and for values of $\kappa_{2}=2.1$ and $\beta=-1$.  As discussed in Sect.~\ref{Num}, when $D_{H}$ is correctly chosen, $c_{N_{4}}(x)$ should be a universal function if the phase has a well-defined (finite) Hausdorff dimension.  We assume $D_{H}=4$ in the rescaling of $c_{N_{4}}(x)$ and $x$ shown in Fig. \ref{VolCorr4K8K12KB1}.  As can be seen in Fig. \ref{VolCorr4K8K12KB1} the peak height of $c_{N_{4}}(x)$ grows with volume, demonstrating that the Hausdorff dimension is not consistent with 4.  Figure \ref{VolCorr4K8K12KB1_v2} shows the rescaled volume-volume correlator $c_{N_{4}}(x)$, this time rescaled with a Hausdorff dimension $D_{H}=12$.  This is the scaling dimension for which the curves give the best overlap.  The large Hausdorff dimension seen in the crinkled region is typical of the collapsed phase.  

\begin{figure}[H] 
  \centering
  \includegraphics[width=0.8\linewidth,natwidth=610,natheight=642]{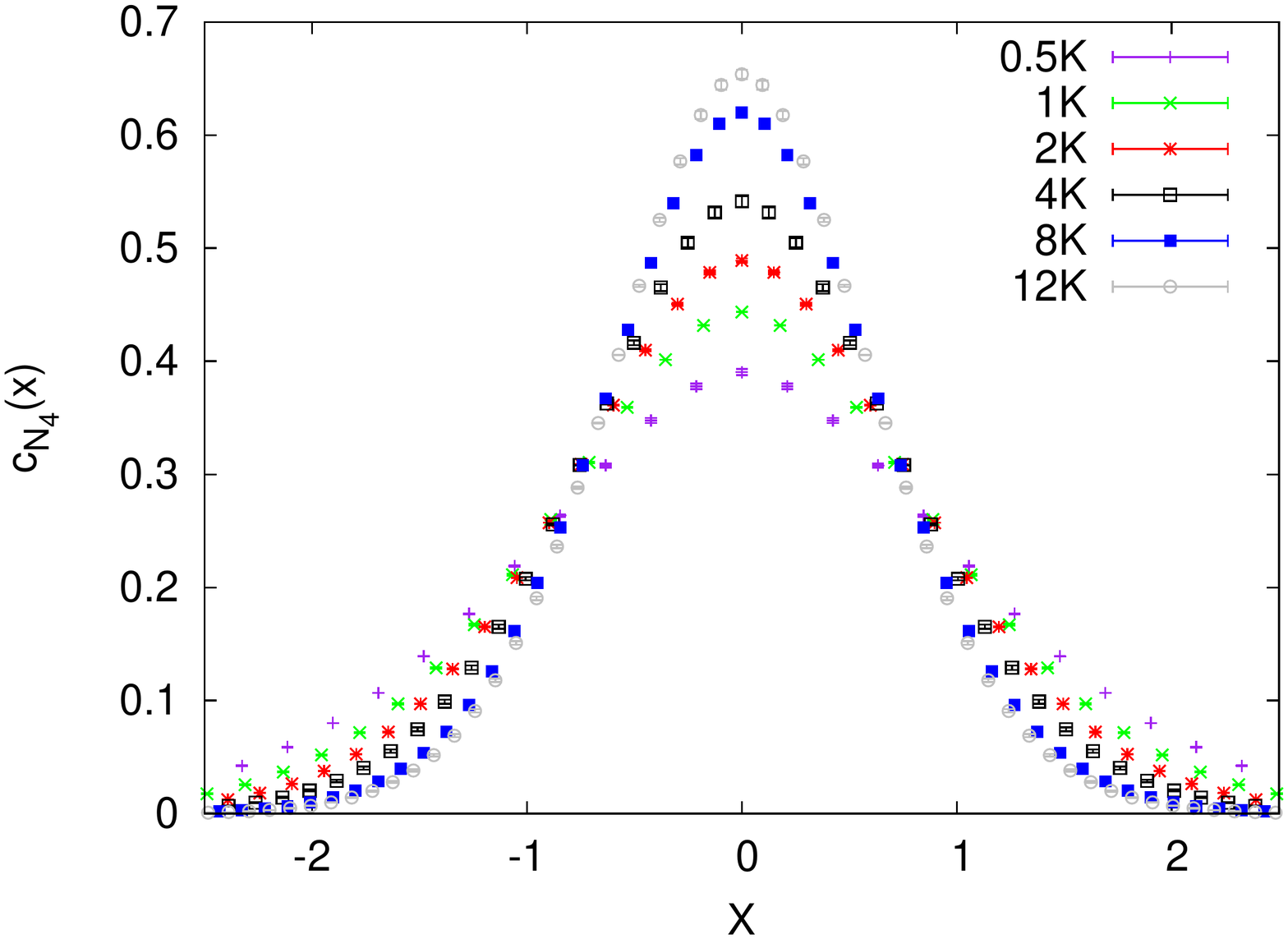}
  \caption{\small The volume-volume correlator as a function of the rescaled variable $x$ for six different lattice volumes within the crinkled region ($\kappa_{2}=2.1$ and $\beta=-1$). $c_{N_{x}}(x)$ is rescaled assuming a Hausdorff dimension $D_{H}=4$. The disagreement between the data indicates that the Hausdorff dimension in the crinkled region is greater than 4.}
\label{VolCorr4K8K12KB1}
\end{figure} 

\begin{figure}[H]
  \centering
  \includegraphics[width=0.8\linewidth,natwidth=610,natheight=642]{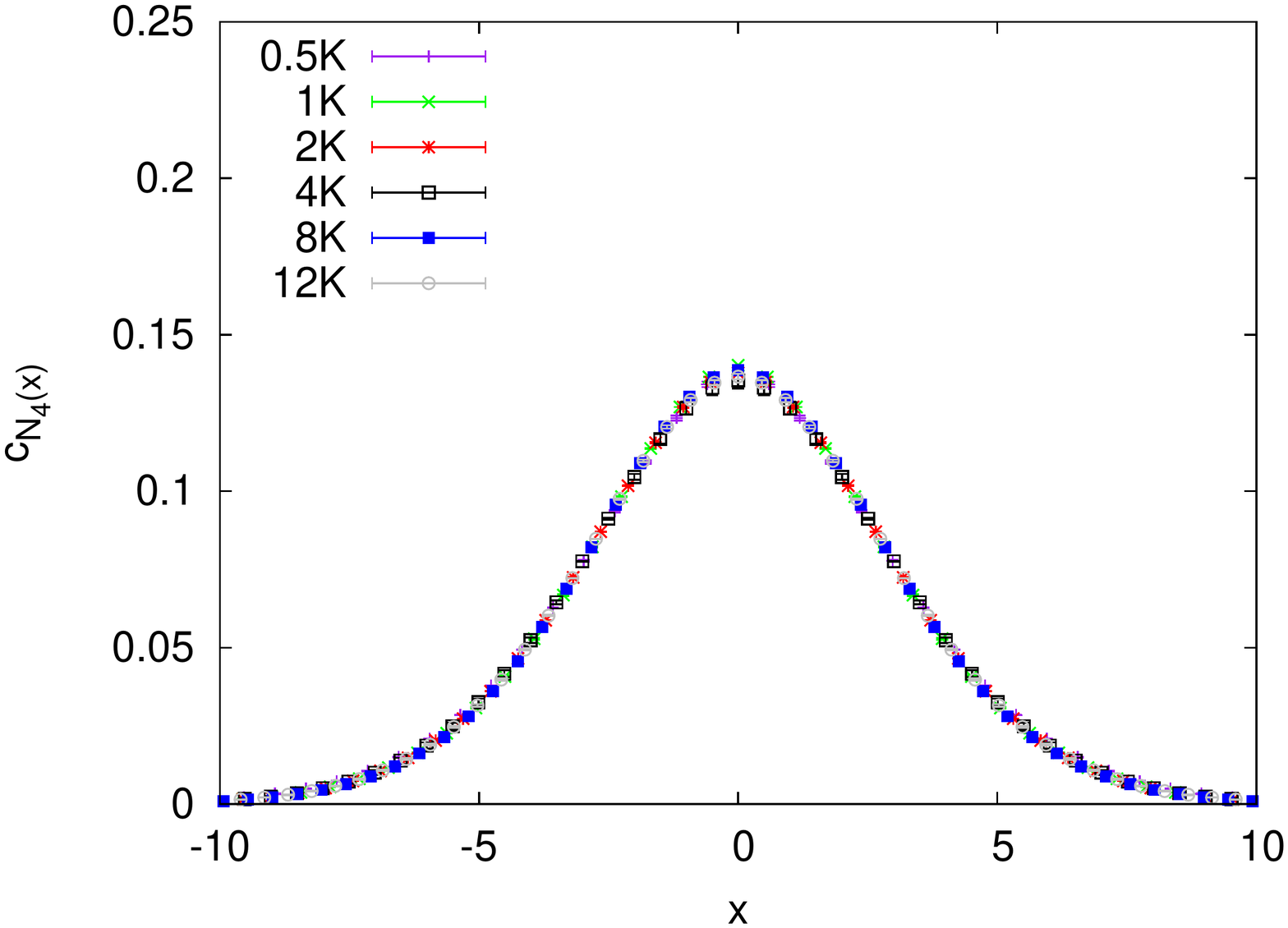}
  \caption{\small The volume-volume correlator as a function of the rescaled variable $x$ for six different lattice volumes within the crinkled region ($\kappa_{2}=2.1$ and $\beta=-1$). $c_{N_{x}}(x)$ is rescaled assuming a Hausdorff dimension $D_{H}=12$.}
\label{VolCorr4K8K12KB1_v2}
\end{figure}

\end{subsection}


\begin{subsection}{The Spectral Dimension}\label{runningspecedt}

Studies of CDT \cite{Ambjorn05} show that the spectral dimension runs as a function of the distance scale probed, with a value around 4 at large distances and a value $\sim 2$ at short distances. Our previous work \cite{Laiho:2011ya} studied the spectral dimension in the crinkled region, where it was tentatively concluded that the spectral dimension approached 4 at long distances and 3/2 at short distances. On this basis we labelled the crinkled region ``the extended phase'', mirroring the notation used by Ref. \cite{Ambjorn05} for a phase with similar properties in CDT. However, the work presented here shows that this region of our phase diagram is not so similar to the extended phase of CDT as first thought. 

We expect finite size effects to be important when $\sigma$ becomes significantly larger than $N_4^{2/D_s}$.  When $\sigma$ is taken large enough, the spectral dimension on a finite lattice volume will decrease as a function of $\sigma$ and eventually approach zero.  Thus, we expect that on our lattice volumes we should be able to take $\sigma$ to a few hundred without seeing the spectral dimension decrease as a function of $\sigma$ due to finite-size effects.  Our calculations are consistent with this expectation, as can be seen in Fig.~\ref{Spec4KB1}.  We have simulations at three different volumes in the crinkled region so that we can explore finite-volume effects in this region explicitly.  In addition to finite-volume effects, we also have to worry about discretization effects.  We expect that discretization errors become large for sufficiently small $\sigma$.  It is possible to get a sense of how small we can take $\sigma$ by looking at the spectral dimension in the branched polymer phase where the answer is a known constant independent of the scale probed, up to discretization effects.  In practice we find that discretization effects are small for the spectral dimension in the branched polymer phase for $\sigma$ above 50-100, depending on the values of the bare lattice parameters.  

We first examine the spectral dimension on the same ensemble used in our previous work. Our new calculation of the spectral dimension on the 4K volume previously presented in Ref. \cite{Laiho:2011ya} now makes use of $\sim$100 times the number of configurations.  We compute the spectral dimension using Eq. (\ref{spec2}) on $\sim$90,000 configurations. The statistical errors are determined from a single elimination jackknife procedure.  In the case of the spectral dimension, we find that binning has little effect on the statistical errors and that autocorrelation effects are small.  Following Ref. \cite{Ambjorn:2005db} we fit the spectral dimension data to the functional form 

\begin{equation}\label{FitFunc}
D_{s}=a-\frac{b}{c+\sigma},
\end{equation} 

\noindent where the constants a, b, and c are free parameters. The fit function, with our new statistics, no longer accurately describes the data, as can be seen by the relatively large $\chi^{2}/ \textrm{d.o.f}$=1.97 that is found when using the full covariance matrix in the estimate of $\chi^{2}$. This translates into a p-value of 
$7\times 10^{-5}$. Although Eq. (\ref{FitFunc}) appears to be a good description of the spectral dimension in CDT, given that our results for the Hausdorff dimension in the crinkled region of EDT differ significantly from the results of CDT, there is no reason to expect Eq.~(\ref{FitFunc}) to be a good description of our data. For the purposes of a comparison between CDT and EDT  it is still interesting to see what a fit to the ansatz Eq.~(\ref{FitFunc}) gives for the asymptotic value of $D_S$ as $\sigma \to \infty$.  It may be that the poor quality of the fit is due to small corrections to Eq.~(\ref{FitFunc}) that the fit is sensitive to because of the small statistical errors in our data.  In the absence of better theoretical guidance as to the correct functional form of $D_S$ in this model, we use Eq.~(\ref{FitFunc}) to see what results are suggested by our data if we assume behavior similar to that of CDT.

 Both correlated and uncorrelated fits give very similar results, as shown in Fig. \ref{Spec4KB1}. A fit to the 4K lattice data with $\sigma$ ranging from 46 to 246 in steps of 4 gives $D_{S}\left(\infty\right)=4.60\pm0.07$. The error here is statistical only.  The systematic error is difficult to estimate given the lack of theoretical guidance on the true asymptotic behavior of $D_S(\sigma)$ in this model, but already we see a hint that $D_S>4$ at large $\sigma$.  For the 8K ensemble we analysed $\sim70,000$ configurations with $\sigma$ ranging from 46 to 246 in steps of 4. Again taking Eq.~(\ref{FitFunc}) as our fit ansatz, we find $D_{S}\left(\infty\right)=5.07\pm0.10$. For the 12K ensemble we analysed $\sim100,000$ configurations with the same procedure and found $D_{S}\left(\infty\right)=5.33\pm0.09$. Based on these values, we conclude that it is very likely that $D_{S}$ at large $\sigma$ is greater than 4 and increases with volume, providing additional evidence that the crinkled region of our model does not have semiclassical behavior like that seen in CDT.

\begin{figure}[H] 
  \centering
  \includegraphics[width=0.8\linewidth,natwidth=610,natheight=642]{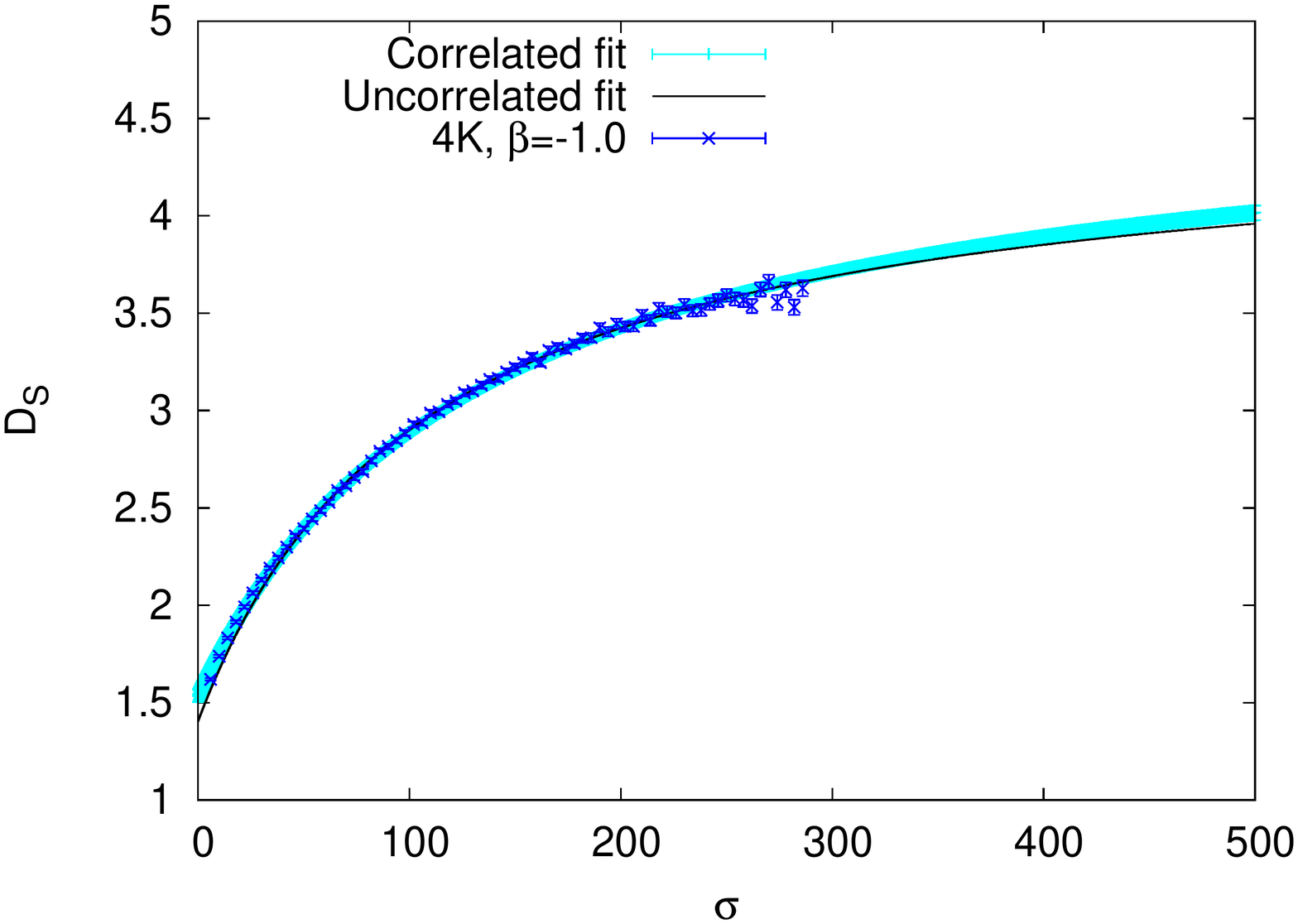}
  \caption{\small The spectral dimension as a function of diffusion time $\sigma$ calculated in the crinkled region ($\kappa_{2}=2.1$, $\beta=-1$) on an ensemble with volume=4,000 4-simplices. An error band is shown for the correlated fit. The uncorrelated fit shows only the central value.}
\label{Spec4KB1}
\end{figure} 

\begin{figure}[H] 
  \centering
  \includegraphics[width=0.8\linewidth,natwidth=610,natheight=642]{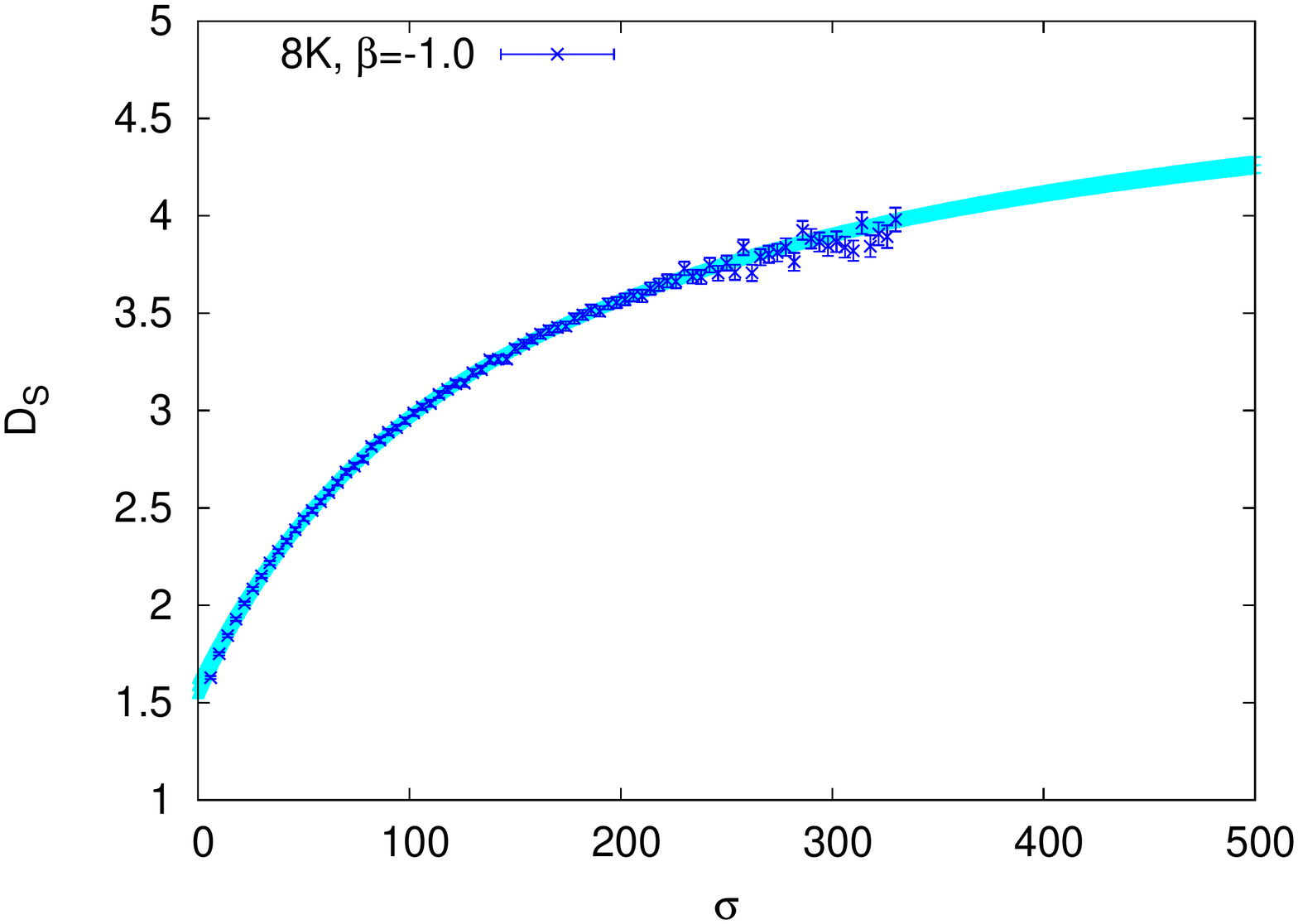}
  \caption{\small The spectral dimension as a function of diffusion time $\sigma$ in the crinkled region ($\kappa_{2}=2.1$, $\beta=-1$) on an ensemble with volume=8,000 4-simplices.}
\label{Spec8KB1}
\end{figure}

\begin{figure}[H] 
  \centering
  \includegraphics[width=0.8\linewidth,natwidth=610,natheight=642]{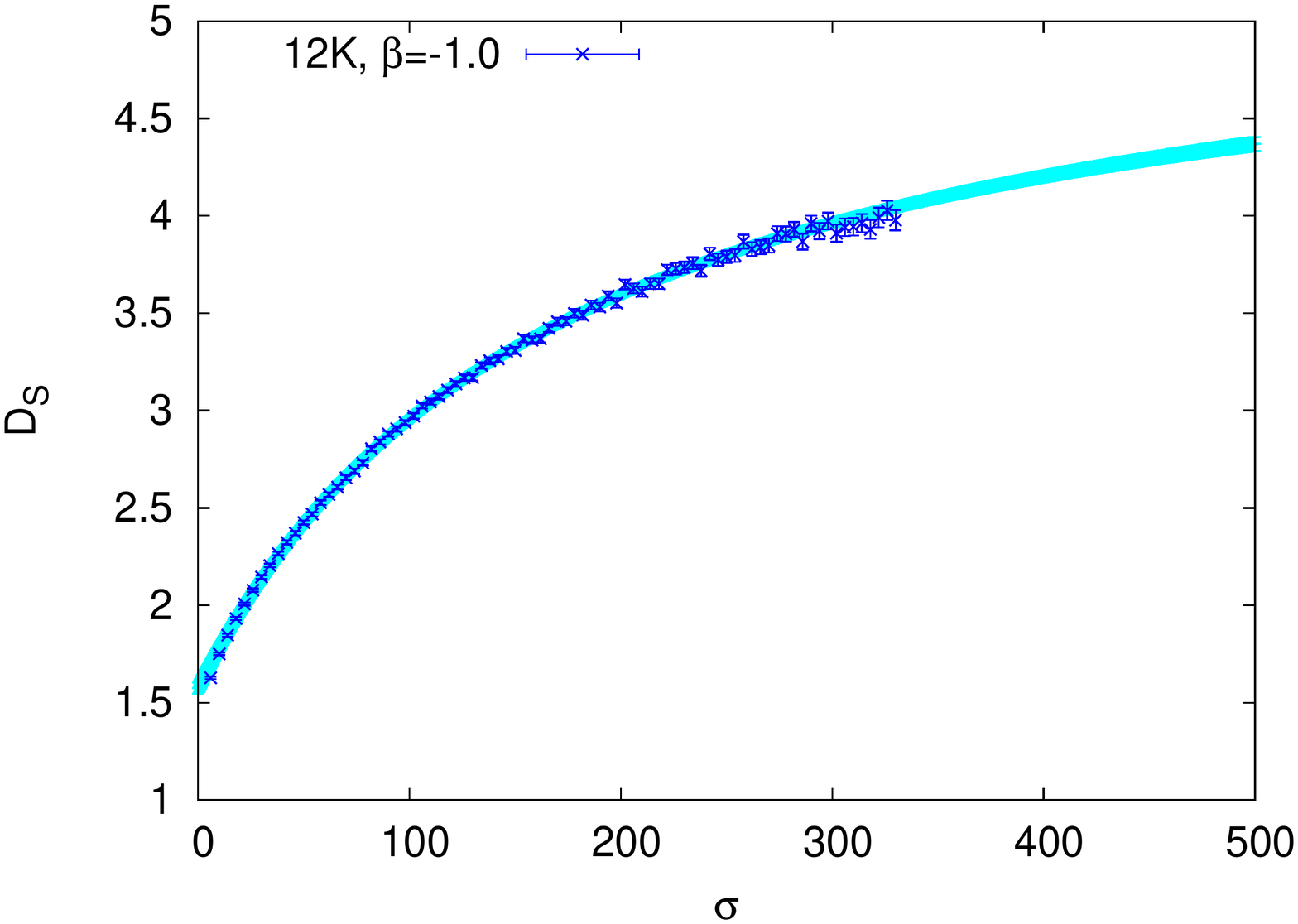}
  \caption{\small The spectral dimension as a function of diffusion time $\sigma$ calculated in the crinkled region ($\kappa_{2}=2.1$, $\beta=-1$) on an ensemble with volume=12,000 4-simplices.}
\label{Spec12KB1}
\end{figure}


\begin{figure}[H] 
  \centering
  \includegraphics[width=0.8\linewidth,natwidth=610,natheight=642]{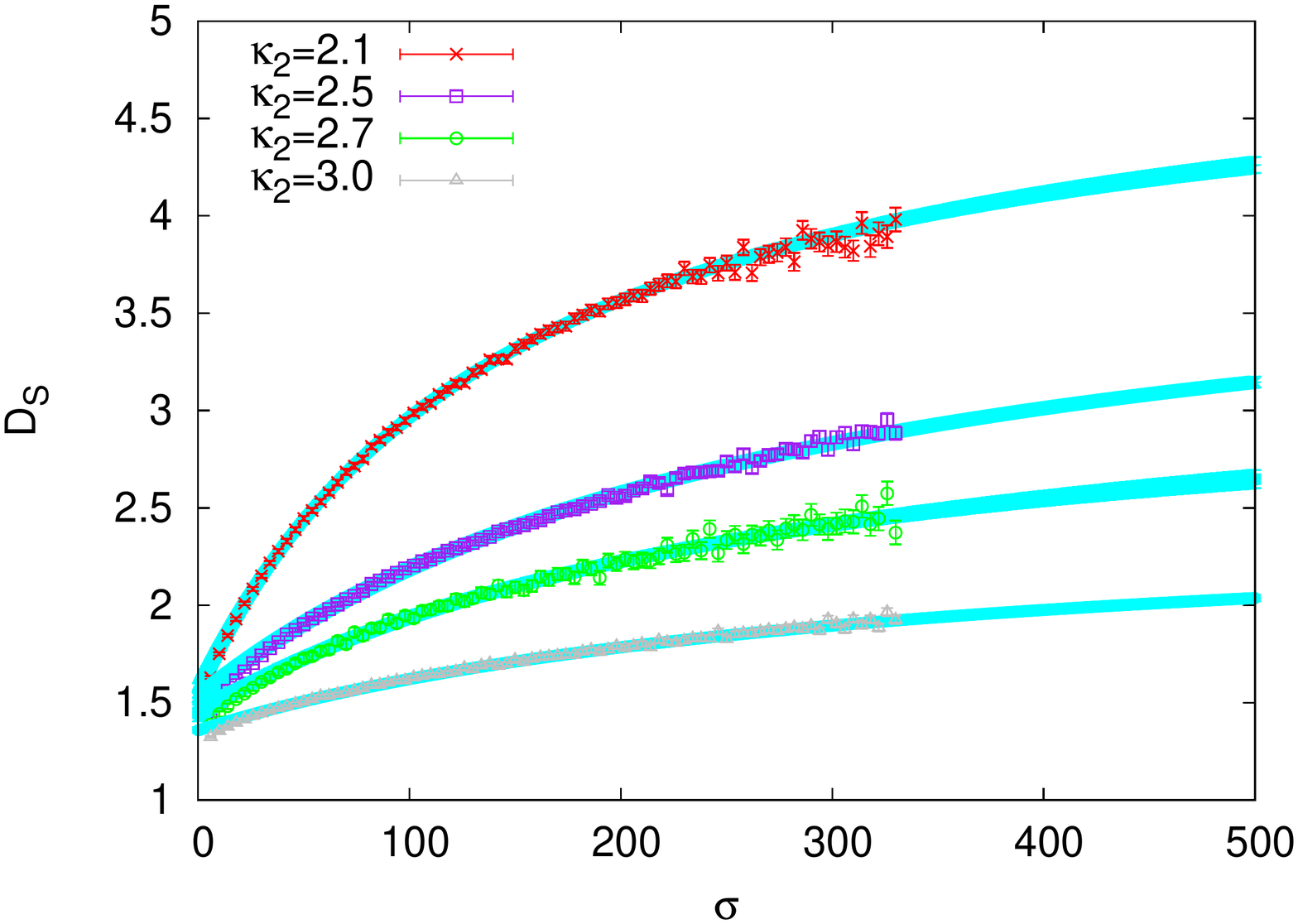}
  \caption{\small The spectral dimension as a function of diffusion time $\sigma$ for four different values of $\kappa_{2}$ with $\beta$ fixed at $-1.0$. All curves in this plot are calculated using a volume of 8000 4-simplices.}
  \label{SpecRel8KB1}
\end{figure}

We also look at $D_{S}\left(\sigma\right)$ as a function of $\kappa_{2}$ for fixed $\beta=-1$ in the crinkled region. Again, we fit the data to Eq.~(\ref{FitFunc}). The range of $\sigma$ values over which the fit function is applied is varied for each $\kappa_{2}$ value. The value of the spectral dimension as $\sigma\rightarrow 0$ for each of these fits is shown in Tab. \ref{SDCTable}. The first error is statistical and the second is a systematic associated with varying the minimum $\sigma$ value used in the fit between 46 and 100.  Again, given the absence of theoretical guidance on the extrapolation, these values should be taken with a grain of salt.  If, for example, $D_S$ is non-monotonic, as seen in renormalization group studies of the truncated effective action \cite{Reuter:2011ah, Rechenberger:2012pm}, then these results could be far off the correct extrapolated value.  

Based on our results, we put forward the following provisional explanation:  in the same way that the scalar curvature smoothly interpolates between collapsed and branched polymer type behavior (see Figs.~\ref{AvgR4K8KB1} and \ref{ChiR4K8KB1}) as $\kappa_{2}$ increases, it is plausible that the spectral dimension does the same. The spectral dimension then approaches the branched polymer value as $\kappa_{2}$ increases, eventually becoming a constant independent of $\sigma$ and taking the value $D_{S}=4/3$. The values of $D_{S}\left(0\right)$ shown in Tab.~\ref{SDCTable} are consistent with this picture.  Given the lack of control over the extrapolation to $\sigma=0$, it is difficult to say more than this.

These results for the spectral dimension indicate that $D_{S}\left(0\right)$ may not be consistent with 3/2 in the crinkled region of the phase diagram of EDT, in contradiction to our original claim in Ref. \cite{Laiho:2011ya}. Thus our results do not resolve the tension between asymptotic safety and black hole entropy scaling. However, the crinkled region has other undesirable features, as we have shown, so the value of the spectral dimension calculated here in what appears to be an unphysical phase is most likely not relevant to the question of what happens in a viable theory of gravity.  A determination of $D_{S}\left(0\right)$ from a lattice approach that realizes Weinberg's asymptotic safety scenario, if indeed this scenario is realized, would be most useful in clarifying this issue.  The recent work of Ref.~\cite{Coumbe:2014noa} finds that the short distance spectral dimension of CDT is compatible with 3/2.  Since CDT has other desirable properties, this lends support to the argument put forth in Ref.~\cite{Laiho:2011ya} that would resolve asymptotic safety and holographic entropy scaling, though much remains to be done before we have the definitive result for $D_S(0)$ from lattice quantum gravity.  Although the present work shows that the crinkled region of EDT does not have the good properties of CDT, we believe our results for the spectral dimension may be of interest for comparison to other approaches where analogs of the "crinkled" or collapsed phase might exist, e.g. in CDT or in renormalization group approaches.  See Refs.~\cite{Shomer:2007vq, Banks:2010tj, Percacci:2010af, Falls:2012nd} for different perspectives on the tension between asymptotic safety and black hole entropy scaling.

\begin{table}[H]
\centering
\caption{A table of the short distance spectral dimension $D_{S}(\sigma\rightarrow 0)$ for four different $\kappa_{2}$ values at $\beta=-1$. $D_{S}(\sigma\rightarrow 0)$ is determined from a fit to the form $a-\frac{b}{c+\sigma}$.  The first error is statistical and the second is a systematic error associated with varying the fit range; the systematic error is not complete because of the absence of a solid theoretical justification for the extrapolation ansatz.}
  \begin{tabular}{|c|c|}
    \hline
    $\kappa_{2}$ & $D_{S}(\sigma\rightarrow 0)$  \\ \hline\hline
    2.1 & $1.52(4)(13)$  \\ \hline
    2.5 & $1.42(3)(23)$  \\ \hline
    2.7 & $1.39(5)(22)$  \\ \hline
    3.0 & $1.355(15)(29)$  \\ \hline
  \end{tabular}
    \label{SDCTable}
\end{table}

\end{subsection}

\end{section}


\begin{section}{Conclusions and Outlook}

This work is the first detailed study of the phase diagram of EDT with a non-trivial measure term using degenerate triangulations.  In this study we have demonstrated that the so-called crinkled region of the phase diagram of EDT with a non-trivial local measure term is not 4-dimensional like the de Sitter phase found in CDT. Results for both the Hausdorff and spectral dimension suggest that the crinkled region of EDT is not a separate phase but a region of the collapsed phase with especially large finite-size effects.    
The phase diagram of EDT with a non-trivial measure term does not have an obvious second-order transition, at least within the parameter values easily accessible to numerical simulations. The transition line separating the collapsed and branched polymer phases is almost certainly first-order, and the boundary between the collapsed phase and the crinkled region appears to be a cross-over. This situation is unlike the results found in the CDT approach, where there is a distinct semiclassical phase, and a second-order transition line separating the semiclassical phase and the collapsed phase. 

We find that although the spectral dimension of EDT in the so-called crinkled region of the phase diagram at small volumes bears a striking resemblance to that of CDT in the semiclassical phase \cite{Laiho:2011ya}, these features do not persist as the volume is increased, and the simulations show behavior that resembles that of the collapsed phase.  In Ref.~\cite{Laiho:2011ya} we found that the short-distance value of the spectral dimension was close to 3/2, and we  argued that this might resolve the tension between asymptotic safety and black hole entropy scaling.  Although the region of the phase diagram studied in Ref.~\cite{Laiho:2011ya} does not look like a good candidate for a semiclassical theory of gravity, this work did in part motivate the study of the short-distance spectral dimension of CDT in Ref.~\cite{Coumbe:2014noa}, where a value for $D_S(0)$ compatible with 3/2 was found, thus providing some support for our initial conjecture in a lattice formulation that does have nice semiclassical properties.  While this result for CDT may be merely a coincidence, we believe it warrants further investigation.

It is still not known whether some modification of the Euclidean approach can reproduce the results of CDT and/or provide a realization of the asymptotic safety scenario. One possible solution is to add higher-order curvature terms to the Euclidean action. In fact it was shown in Ref.~\cite{Stelle:1976gc} that adding $R^{2}$ terms leads to a perturbatively renormalizable theory of gravity, albeit one that is expected to be non-unitary when Wick rotated back to Lorentzian signature. The addition of an $R^{2}$ term to the Einstein-Hilbert action in the EDT formulation was investigated in Ref.~\cite{Ambjorn:1992aw}, where the $R^{2}$ term was discretized in a straightforward way. The resulting phase diagram was not essentially different from the phase diagram found in this work, though it may be worth revisiting it with modern computing resources. It may also be worth investigating the addition of a second independent $R^{2}$ term, since there are two independent $R^{2}$ terms (the third is a topological invariant in 4-dimensions) that are marginal by power counting. 

\end{section}

\section*{Acknowledgments}

We thank W. Bardeen, S. Catterall, D. Litim, D. Miller, F. Saueressig, and C. White for useful discussions. We also thank Mike Czajkowski and Jayanth Neelakanta for their assistance in precisely locating the first order phase transition using multihistogram methods. This work was funded by STFC, the Scottish Universities Physics Alliance, and Syracuse University.  Computing was done on the Darwin Supercomputer as part of STFC's DiRAC facility jointly funded by STFC, BIS and the Universities of Cambridge and Glasgow. D.N.C acknowledges the support of the grant DEC-2012/06/A/ST2/00389 from the National Science Centre Poland.


\bibliographystyle{unsrt}
\bibliography{Master}



\end{document}